\documentclass[11pt,a4paper]{article}

\usepackage[latin1]{inputenc}
\usepackage[english]{babel}
\usepackage{amsfonts}
\usepackage{amsmath}    %NOTE: \beq, \eeq abbrev. won't work
\usepackage{amssymb}
\usepackage{amsthm}
\usepackage[pdftex]{color}
\usepackage{graphicx}

\usepackage[pdftex,colorlinks, pdfpagelabels]{hyperref}
\usepackage[figure,table]{hypcap} % Correct a problem with hyperref
\hypersetup{
   bookmarksnumbered,
   pdfstartview={FitV},
   pdfpagemode={UseOutlines},
   pdfauthor={Jasper Hasenkamp, Joern Kersten},
   pdftitle={Leptogenesis, Gravitino Dark Matter and Entropy Production},
   pdfsubject={}
   ,citecolor={blue},
   linkcolor={blue},
   urlcolor={blue},
   filecolor={blue}
} 

\def\a{\alpha}

\def\k{\kappa}

\def\D{\Delta}

\def\G{\Gamma}

\def\O{\Omega}

\newcommand{\meV}{\text{ meV}}
\newcommand{\eV}{\text{ eV}}

\newcommand{\MeV}{\text{ MeV}}
\newcommand{\GeV}{\text{ GeV}}
\newcommand{\TeV}{\text{ TeV}}
\newcommand{\Mpc}{\text{ Mpc}}
\newcommand{\km}{\text{ km}}
\newcommand{\seconds}{\text{ s}}

\newcommand{\mgrav}{m_{3/2}}
\newcommand{\gr}{\ensuremath{\Psi_{3/2}}}
\newcommand{\mplanck}{\ensuremath{M_{\text{Pl}}}}
\newcommand{\Mr}{\ensuremath{M_{\nu_\text{R}^1}}}
\newcommand{\meff}{\ensuremath{\widetilde m_{\nu_\text{L}^1}}}

\newcommand{\CiteSeeSaw}{\cite{Minkowski:1977sc,Yanagida:1980,Glashow:1979vf,Gell-Mann:1980vs,Mohapatra:1980ia}}

\begin{document}
 
\date{\mbox{ }}

\title{ 
{\normalsize     
%\bf \hfill\mbox{}\\
10th August 2010 \hfill\mbox{}\\}
\vspace{2cm}
\bf Leptogenesis, Gravitino Dark Matter and Entropy Production \\[8mm]}
%
%\vspace{2cm} 
\author{Jasper Hasenkamp and J\"{o}rn Kersten\\[2mm]
{\small\it II.\ Institute for Theoretical Physics, University of Hamburg, Germany}\\
{\small\tt Jasper.Hasenkamp,Joern.Kersten@desy.de}
}
\maketitle

\thispagestyle{empty}

\vspace{1cm}
\begin{abstract}
\noindent
Many extensions of the Standard Model predict super-weakly interacting
particles, which typically have to decay before Big Bang Nucleosynthesis
(BBN).
The entropy produced in the decays may help to reconcile
thermal leptogenesis and BBN in scenarios with 
gravitino dark matter, which is usually difficult due to 
late decays of the next-to-lightest supersymmetric particle (NLSP)
spoiling the predictions of BBN\@.
We study this possibility for a general neutralino NLSP\@.

We elaborate general properties of the scenario and strong
constraints on the entropy-producing particle. 
As an example, we consider the saxion from the
axion multiplet and show that, while enabling
a solution of the strong CP problem, it can also produce a
suitable amount of entropy.
\end{abstract}

\newpage

\section{Introduction}

Thermal leptogenesis \cite{Fukugita:1986hr} is an attractive mechanism
for generating the baryon asymmetry of the Universe, since it requires no
additional ingredients beyond the see-saw scenario \CiteSeeSaw\
introduced to explain the smallness of the observed neutrino masses.
However, in supergravity theories the required large reheating
temperature results in a copious production of
gravitinos~\cite{Ellis:1982yb,Nanopoulos:1983up}.
If the gravitino is heavier than other superparticles, this is
problematic since it typically decays during or after 
Big Bang Nucleosynthesis (BBN) due to its extremely
 weak interactions.  The energetic decay products can then cause
unacceptable changes of the primordial light element
abundances~\cite{Falomkin:1984eu,Ellis:1984eq}.  If on the other hand the gravitino
is the lightest supersymmetric particle (LSP), 
high temperatures tend to lead to a relic gravitino density
exceeding the observed dark matter density~\cite{Khlopov:1984pf}.  
In addition, the next-to-lightest supersymmetric particle (NLSP)
 becomes long-lived and thus can cause 
similar problems with
BBN as an unstable gravitino~\cite{Moroi:1993mb}.

Apart from abandoning supersymmetry or thermal leptogenesis, the
gravitino problem has two further solutions.  The first one is to reconcile
thermal leptogenesis with a smaller reheating temperature.  This is
possible if there is a resonant enhancement of the generated
asymmetry~\cite{Flanz:1996fb,Pilaftsis:1997jf,Covi:1996wh},
some fine-tuning that violates the naturalness assumptions entering
into the lower bound on the reheating temperature~\cite{Raidal:2004vt},
or a violation of R-parity~\cite{Farzan:2005ez}.

The second option is to accept a large reheating temperature and
tackle the problems associated with the gravitino.
One possibility is a very heavy gravitino that decays before BBN~\cite{Weinberg:1982zq}.
Alternatively, the gravitino could be very light, thus both avoiding a
too large relic density and letting the NLSP decay before BBN~\cite{Moroi:1993mb}.
The relic density of a gravitino LSP with mass in the GeV range can also
be acceptable for rather large reheating
temperatures~\cite{Bolz:1998ek}, but in this case we have to protect BBN
from the NLSP decays.
This is possible if the NLSP decays relatively fast due to 
R-parity violation~\cite{Buchmuller:2007ui} or additional decay modes into hidden sector states~\cite{DeSimone:2010tr,Cheung:2010qf},
 if its decay products are
only weakly interacting~\cite{Fujii:2003nr}
or very low-energetic~\cite{Boubekeur:2010nt} and therefore harmless,
or if its abundance is exceptionally small.  The last alternative can
occur for a stau NLSP~\cite{Ratz:2008qh,Pradler:2008qc} 
in exceptional regions of parameter space, 
or for any NLSP whose abundance is diluted by entropy produced in late
decays of another particle~\cite{Buchmuller:2006tt,Pradler:2006hh,Kasuya:2007cy}.
Of course, a combination of the various options is
possible~\cite{Asaka:2000ew,Baltz:2001rq,Fujii:2002fv,Fujii:2003iw}.

In this work, we study a scenario where the gravitino is the LSP and
forms the dark matter. The abundance of the NLSP is
decreased by late-time entropy production.  
This dilution has the additional motivation that long-lived particles
are anyway present in many extensions of the Standard Model (SM) and usually have to
decay before BBN to avoid problems, producing entropy in the process.
The solution of the gravitino problem may thus be viewed as a
complimentary by-product of such an extension rather than an additional
complication of the Minimal Supersymmetric Standard Model (MSSM)\@. Of the several candidates for the NLSP,
we consider the lightest neutralino.

The paper is organised as follows. In Sec.~\ref{sec:TL+GDM}, we review briefly
the scenario of thermal leptogenesis with gravitino dark matter.
Sec.~\ref{sec:entropy} discusses the impact of
entropy production on this scenario and the mechanism of 
entropy production by decaying matter.
In Sec.~\ref{sec:nlsp}, we present BBN constraints on a
general neutralino NLSP after suitable dilution.
We elaborate general
properties of the scenario and strong
constraints on the entropy-producing particle
in Sec.~\ref{sec:candidate}, where we also
consider as an example the saxion from the 
axion multiplet. 

\section{Leptogenesis and Gravitino Dark Matter\dots}
\label{sec:TL+GDM}
In baryogenesis via standard thermal leptogenesis a cosmic lepton asymmetry is generated by CP-violating out-of-equilibrium decays of heavy right-handed Majorana neutrinos $\nu_\text{R}^i$. 
Non-perturbative sphaleron processes~\cite{Kuzmin:1985mm,Bodeker:1999gx} convert
the lepton asymmetry into a baryon asymmetry $\eta_\text{B}$.  
 In the case of hierarchical masses the maximal resulting baryon-to-photon ratio of the Universe can be given as~\cite{Buchmuller:2002rq,Davidson:2008bu}
\begin{equation}
\label{eq:etab}
\eta_\text{B}^\text{max} \simeq 9.6 \times 10^{-10} \, \D^{-1} 
\left(\frac{M_{\nu_\text{R}^1}}{2 \times 10^{9} \GeV}\right) 
\left(\frac{m_{\nu_\text{L}^3}}{0.05\eV}\right)
\left(\frac{\k_0}{0.18}\right)
\end{equation}
for the MSSM and weak washout,
while the observed value lies in the range $5.89 \times 10^{-10} < \eta_\text{B}^\text{obs} < 6.49 \times 10^{-10}$~($2\sigma$)~\cite{Komatsu:2010fb}.
$M_{\nu_\text{R}^1}$ denotes the Majorana mass of the lightest right-handed neutrino.
The given baryon asymmetry is maximal in the sense that the CP violation in the decays is chosen to be maximal~\cite{Davidson:2002qv}. 
Since thermal leptogenesis strongly favours
hierarchical light neutrino masses~\cite{Buchmuller:2003gz}, the mass $m_{\nu_\text{L}^3}$ of the heaviest left-handed neutrino has to be close to $\sqrt{\Delta m^2_{31}} \simeq 0.050\eV$, using the best-fit value from neutrino data~\cite{GonzalezGarcia:2010er}
and assuming a normal mass ordering.
The efficiency factor $\k_0$ should be computed case-by-case
by solving the relevant Boltzmann equations~\cite{Barbieri:1999ma,Plumacher:1996kc,Plumacher:1997ru}.
For zero initial $\nu_\text{R}^1$ abundance in the small $M_{\nu_\text{R}^1}$ regime~\cite{Buchmuller:2002rq}, 
i.e.\ for $M_{\nu_\text{R}^1}\lesssim 4 \times 10^{13} \GeV$,
the maximal value is $\k_0^\text{peak} \simeq 0.18$~\cite{Buchmuller:2004nz}.
This value 
is reached for
 \begin{equation}
  \meff \simeq m_\ast= \frac{ 8\pi^2 \sqrt{g_\ast} }{3  \sqrt{10}} \frac{v^2}{M_\text{Pl}} \simeq  1.6 \times 10^{-3}\eV \, ,
 \end{equation}
where $m_\ast$ is known as the equilibrium neutrino mass, $v \simeq 174\GeV$, and
$M_\text{Pl} \simeq 2.44 \times 10^{18}\GeV$.
The effective neutrino mass 
\begin{equation}
 \meff = \frac{\bigl( m_\text{D}^\dagger m_\text{D}^{} \bigr)_{11}}{\Mr}
\end{equation}
equals the mass of the lightest neutrino if the Dirac mass matrix $m_\text{D}$ is diagonal. Its natural range is 
$m_{\nu_\text{L}^1} < \widetilde m_{\nu_\text{L}^1} < m_{\nu_\text{L}^3}$.
The parameter $\D$ denotes the dilution factor by entropy production after the  decay
 of the right-handed neutrinos. It equals one in standard cosmology, while we will consider the general case $\D \geq 1$ later on.

There are some uncertainties entering~\eqref{eq:etab}.
Possible spectator field uncertainties~\cite{Buchmuller:2001sr} 
and flavour effects~\cite{Abada:2006fw,Nardi:2006fx} are neglected, 
and the naive sphaleron conversion factor~\cite{Khlebnikov:1988sr,Harvey:1990qw} is used.
We have assumed the particle content of
the MSSM with $g_\ast = 228.75$ for the number of effectively massless
degrees of freedom at high temperatures. To be conservative
we consider the effects of the MSSM by a factor $2 \sqrt{2}$ 
relative to the SM, which is valid for weak washout~\cite{Davidson:2008bu}.
For strong washout this factor reduces to $\sqrt{2}$.

We see from~\eqref{eq:etab} that leptogenesis in its minimal version as described
above can generate the observed baryon-to-photon ratio of the Universe,
because $\eta_\text{B}^\text{max}$ can exceed $\eta_\text{B}^\text{obs}$.
On the other hand, it is clear that there is a lower bound $M_{\nu_\text{R}^1} \gtrsim 2 \times 10^{9} \GeV$. 

It is especially appealing for the considered neutrino mass range that leptogenesis can emerge as the unique source
of the cosmological baryon asymmetry~\cite{Buchmuller:2003gz}. 
Wash-out processes may reduce a pre-existing asymmetry
by two to
three orders of magnitude for the situation of~\eqref{eq:etab}.
Stronger washout decreases the efficiency factor and thus requires a
larger right-handed neutrino mass to keep $\eta_\text{B}^\text{max} \geq \eta_\text{B}^\text{obs}$.
For thermal leptogenesis, the bound on the lightest right-handed
neutrino mass can be translated into a lower bound on the reheating
temperature after inflation, $T_\text{R} \gtrsim M_{\nu_\text{R}^1}$. 
In the strong washout regime, i.e.\ for $\meff > m_\ast$,
this changes to $T_\text{R} \gtrsim 0.1 \, \Mr$~\cite{Buchmuller:2004nz},
but we cannot relax the bound on the absolute value of $T_\text{R}$, since in
this case the efficiency factor decreases as well, requiring a larger
\Mr.

The required high temperatures also lead to thermal production of a
significant gravitino relic density
\begin{equation}
\label{eq:otp}
 \O_{3/2}^\text{tp} h^2 = \mgrav \, Y_{3/2}^\text{tp}(T_0) \, \frac{s(T_0) \, h^2}{\rho_0} \, ,
\end{equation}
where $s(T_0)$ refers to today's entropy density of the Universe.
Together with the Hubble constant $h$ in units of 
$100 \km \Mpc^{-1} \seconds^{-1}$ and today's critical density $\rho_0$, we
obtain $s(T_0) h^2 / \rho_0 \simeq 2.8 \times 10^{8} \GeV^{-1}$.
The gravitino abundance for low temperatures $T_{\text{low}} \ll T_\text{R}$
is given by~\cite{Bolz:2000fu,Pradler:2006qh}
\begin{equation}
 \label{eq:ytp}
 Y_{3/2}^\text{tp}(T_{\text{low}}) \simeq \sum_{i=1}^3 y_i g_i^2(T_\text{R}) \left( 1+ \frac{M_i^2(T_\text{R})}{3 \mgrav^2} \right)
 \ln{\left(\frac{k_i}{g_i(T_\text{R})}\right)} \left(\frac{T_\text{R}}{10^{10} \GeV}\right) ,
\end{equation}
where the gauge couplings $g_i = (g', g, g_s)$, 
the gaugino mass parameters $M_i$ as well as the constants 
$k_i = (1.266, 1.312, 1.271)$ and
$y_i/10^{-12} = (0.653, 1.604, 4.276)$
are associated with the gauge groups
U(1)$_\text{Y}$, SU(2)$_\text{L}$ and SU(3)$_\text{C}$,
respectively.

Without entropy production, the gravitino yield from thermal production
at the present temperature
$Y_{3/2}^\text{tp}(T_0) = Y_{3/2}^\text{tp}(T_{\text{low}})$. With entropy
production after the gravitino production in the early Universe,
\begin{equation}
\label{eq:ydelta}
 Y_{3/2}^\text{tp}(T_0) = \D^{-1} \, Y_{3/2}^\text{tp}(T_{\text{low}}) \, .
\end{equation}
As  mentioned before, $\D=1$ in standard cosmology. For reasons
that we will explain below, we will consider 
late-time entropy production at $T \ll M_{\nu_\text{R}^1},\,T_{\text{low}}$.
Then the same dilution factor $\D$ appears in \eqref{eq:etab} and~\eqref{eq:ydelta}.%
\footnote{For $T_\text{R} \gg M_{\nu_\text{R}^1}$, it could be possible to produce
entropy in between, so that only $\Omega_{3/2}$ would be diluted but not
$\eta_\text{B}$.}

From~\eqref{eq:otp} and~\eqref{eq:ytp} we see that since the
gravitino is the LSP, for fixed gaugino masses the relic gravitino
density typically decreases for increasing gravitino mass. 
Assuming universal gaugino masses at the GUT scale, we can approximate
\begin{equation}
\label{eq:o32est}
 \O_{3/2}^\text{tp} h^2 \simeq 0.11 \, \D^{-1} 
 \left( \frac{T_\text{R}}{3 \times 10^8 \GeV }\right) 
 \left( \frac{M_{\widetilde g}(m_\text{Z})}{10^3 \GeV}\right)^2 
 \left( \frac{10 \GeV}{\mgrav}\right) .
\end{equation}
Thus, for given reheating temperature and gaugino masses, we obtain a lower bound on the gravitino mass exploiting the requirement 
$\O_{3/2} h^2 \leq \O_\text{DM} h^2 = 0.112 \pm 0.007$
($2\sigma$)~\cite{Komatsu:2010fb}. 

In order to summarise the issues discussed so far, we combine
\eqref{eq:etab} and \eqref{eq:o32est} using the best-case relation $T_\text{R}
\simeq M_{\nu_\text{R}^1}$ to eliminate the right-handed neutrino mass, arriving at
\begin{multline}
\label{eq:etab2}
 \eta_\text{B}^\text{max} \simeq 1.4 \times 10^{-10}
 \left( \frac{\O_{3/2}^\text{tp} h^2}{0.11} \right) \left( \frac{10^3 \GeV}{M_{\widetilde g}(m_\text{Z})} \right)^2 \left( \frac{\mgrav}{10 \GeV} \right)
 \\
 {}\times \left(\frac{m_{\nu_\text{L}^3}}{0.05\eV}\right) \left(\frac{\k_0}{0.18}\right) .
\end{multline}
Note that the dilution factor $\D$ cancels out. 
Recalling the discussion after~\eqref{eq:etab}, $m_{\nu_\text{L}^3}$ cannot be raised without lowering $\k_0$.
Thus, even for the most optimistic scenario with $T_\text{R}=2 \times 10^9 \GeV$
the gravitino mass is restricted to a rather large value $\gtrsim 40 \GeV$.
In other words, there is considerable tension between thermal
leptogenesis and gravitino dark matter.

Even worse, the NLSP decay problem is a definite clash between both
notions, since gravitino LSP masses larger than about $10\GeV$ are
excluded in most cases.
In the MSSM with conserved R-parity,
the NLSP has to decay into the gravitino and SM particles. It decays typically with a long lifetime due to the extremely weak interactions of the gravitino. If these decays occur during or
after BBN, the emitted SM particles can change the
primordial abundances of the light 
elements~\cite{Moroi:1993mb,Jedamzik:2006xz,Kawasaki:2008qe}.
Specific setups like the Constrained MSSM have been studied in~\cite{Ellis:2003dn,Roszkowski:2004jd,Ellis:2005ii,Cerdeno:2005eu,Bailly:2009pe}. 
Only in exceptional regions of the parameter
space~\cite{Boubekeur:2010nt,Ratz:2008qh,Pradler:2008qc},
a stau NLSP not orders of magnitude heavier than the gravitino could be
consistent with all constraints, allowing reheating temperatures $T_\text{R} \sim 10^9 \GeV$.
Generically a conservative upper bound on the reheating temperature $T_\text{R} \lesssim \text{ few } 10^8 \GeV$ is found.
Since the maximally produced baryon asymmetry is too small,
various NLSP candidates have been investigated more model-independently~\cite{Feng:2004mt,Kawasaki:2008qe} to
identify best-case scenarios: 
a sneutrino~\cite{Fujii:2003nr,Buchmuller:2006nx,Kanzaki:2006hm,Covi:2007xj,Ellis:2008as,Olechowski:2009bd}
 actually could allow large enough reheating temperatures
with reasonable masses due to its invisible decays.
A stop~\cite{Berger:2008ti,Kusakabe:2009jt} or a
general neutralino~\cite{Covi:2009bk} are not reconcilable with thermal leptogenesis for masses below a TeV\@.
They would definitely require $\mgrav < 10 \GeV$.

Altogether, the strongest conflict between thermal leptogenesis and gravitino dark matter is found in the NLSP decay
problem. This is embodied in~\eqref{eq:etab2} by the restriction to small gravitino masses, $\mgrav \leq 10 \GeV$. 

Entropy production after the freeze-out of the NLSP dilutes its density.
Thus, late-time entropy production can naively
resolve this conflict for any NLSP within or without a specific model. The relic density prior to its decay,
\begin{equation}
 \O_\text{nlsp} = \D^{-1} \, \O_\text{nlsp}^\text{fo} \, ,
\end{equation}
is reduced compared to its freeze-out density $\O_\text{nlsp}^\text{fo}$ by the dilution factor $\D$, which is the same as in~\eqref{eq:etab} and~\eqref{eq:ydelta}.
In Sec.~\ref{sec:nlsp} we show how BBN constraints on a general neutralino with a gravitino LSP with
$\mgrav = 100 \GeV$ are softened by $\D > 1$.

\section{\dots{}with late-time Entropy Production}
\label{sec:entropy}

\subsection{Thermal Leptogenesis and Gravitino Yield with late-time Entropy Production}
From~\eqref{eq:etab} we see that a significant dilution $\D > 1$ can
only be compensated by a larger $M_{\nu_\text{R}^1}$, since all the other
parameters are chosen already to be optimal.
Due to the requirement $T_\text{R} \gtrsim M_{\nu_\text{R}^1}$ this gives a linear shift of the 
required reheating temperature.
Since the gravitino density depends also linearly on the reheating temperature~\eqref{eq:ytp}
and is diluted in the same way as the baryon density,
such a compensation seems to give a trivial shift of the problem to higher
reheating temperatures.  However, there are aspects that do not show up
in~\eqref{eq:etab} and~\eqref{eq:etab2}, in addition to the impact on
the NLSP decay problem.

Most importantly, 
in the domain of large $\Mr$ washout processes reduce the efficiency
factor $\k_0$ exponentially.  In the case of hierarchical neutrinos,
this domain corresponds to $\Mr > 4 \times 10^{13} \GeV$.  From this we
would obtain $\D < 2 \times 10^4$.  However,
while at low $\Mr$ many numeric examples and an analytic approximation
for $\k_0$~\cite{Buchmuller:2004nz} exist in the literature, the
situation for larger $\Mr$ is less well-studied.  As an additional
complication, for $T \gg 10^9 \GeV$ more spectator processes are in
equilibrium and thus should be taken into account.  Hence, it is
not clear whether the maximal value $\kappa_0^\text{peak}$
can be reached for $\Mr \sim 4 \times 10^{13} \GeV$.
Consequently, we expect an upper bound 
\begin{equation}
\label{eq:dboundfromTL}
 \D < \D^\text{max} \sim 10^3 \,\dots\, 10^4 \, .
\end{equation}
We would like to stress that this is an intrinsic bound of the problem.
It is stronger than bounds from perturbativity of Yukawa couplings ($\D < 10^5$) or 
the requirement of a reheating temperature below the GUT scale ($\D < 10^7$).

We remark that according to Fig.~6b of~\cite{Buchmuller:2002rq} there is
a much stronger bound with roughly $\D^\text{max} < 10^2$ for quasi-degenerate
neutrino masses.  Thus, thermal leptogenesis with late-time entropy
production requires hierarchical neutrinos even more than thermal
leptogenesis already does.

Late-time entropy production leads to a strong reduction of the allowed parameter space
for successful thermal leptogenesis. Since the required minimal $\Mr$ is increased,
the range of allowed values for $\k_0$ and the neutrino mass parameters is reduced. 
However,
the same region of parameter space is already favoured by the
need to keep the reheating temperature as low as possible in order to
avoid the overproduction of gravitinos.
Therefore, late-time entropy production does not reduce the parameter space of
thermal leptogenesis with gravitino dark matter.

In~\eqref{eq:ytp} one has to consider the impact of the running couplings and masses due to the shift of the reheating
temperature.  For example, if we increase $T_\text{R}$ from $3 \times 10^9\GeV$
to $3 \times 10^{13}\GeV$ and choose $\Delta=10^4$ to compensate,
$\O_{3/2}^\text{tp}$ decreases by $25\%$.%
\footnote{Besides, the electroweak contributions double their
contribution to the total yield to about $30\%$.}
Note that this effect is unavoidable and softens the tension between
thermal leptogenesis and gravitino dark matter already before
considering the impact of entropy production on the NLSP decay problem.

Another possibility is a gravitino
with such a small mass that it comes into thermal equilibrium after
reheating.  Then its relic abundance becomes
independent of the reheating temperature, which allows $T_\text{R} \gg \Mr$.
Taking into account the lower limit on the mass of a warm dark
matter particle~\cite{Viel:2005qj}, it turns out that its relic energy
density exceeds the observed dark matter density in standard cosmology.
However, already a $\D$ of a few 
dilutes the gravitino sufficiently to make it
viable warm dark matter again~\cite{Baltz:2001rq,Fujii:2002fv}.
For $\D \simeq 10^3$ it forms cold dark matter
with $\mgrav \simeq 1 \MeV$~\cite{Fujii:2003iw,Lemoine:2005hu,Jedamzik:2005ir}.
Note that for these small masses the NLSP decays 
before BBN, so that the decay problem is absent.

\subsection{Entropy Production by decaying Matter}
\label{sec:EntropyFromDecays}
In this section we discuss briefly how decaying matter can produce considerable entropy
in the early Universe~\cite{Scherrer:1984fd,Kolb:1990vq}. We consider a non-relativistic and long-lived particle
species $\phi$ with chemical potential $\mu =0$ in a flat Friedmann-Robertson-Walker Universe.
When $\phi$ drops out of chemical equilibrium, its
abundance $Y_\phi=n_\phi/s$ ``freezes out'', where $n_\phi$ denotes its number density and 
\begin{equation}
\label{eq:s}
 s= \frac{2 \pi^2}{45} \, g_{\ast}(T) \, T^3
\end{equation}
the entropy density of the Universe.%
\footnote{For simplicity we use $g_\ast$ only, 
since the temperatures occurring in this work are above $1 \MeV$, where $g_{\ast S}=g_\ast$.}
 $Y_\phi$ could also be generated from inflaton
decay or thermally after reheating, if $\phi$ never enters chemical equilibrium. The contribution
of non-relativistic particles to the energy density decreases as
$\rho_\text{mat} \propto R^{-3}$, where
$R$ denotes the scale factor. Since 
the energy density of radiation in the Universe,
\begin{equation}
\label{eq:rhorad}
 \rho_\text{rad}=\frac{\pi^2}{30} \, g_\ast(T) \, T^4 \, ,
\end{equation}
decreases $\propto R^{-4}$, $\rho_\text{mat}/\rho_\text{rad}$ grows $\propto R$. 
Since $R$ grows with time, at some time $t^{=}_\phi$ or temperature $T^{=}_\phi$ the unstable species $\phi$ comes to dominate
the energy density automatically,
 if its lifetime $\tau_\phi > t^{=}_\phi$.
If  the Universe has been dominated by radiation before, it enters a
phase of matter domination that lasts roughly till
$\phi$ exponentially decays at $\tau_\phi$. Here we assume that everything released by the particle decay is rapidly
thermalized, i.e.\
on timescales $\D t \ll H^{-1} \simeq$ expansion time.
At some intermediate time
$t \simeq t^{=}_\phi \, (\tau_\phi / t^{=}_\phi)^{3/5}$ the radiation produced in decays
of $\phi$ starts to become the dominant component of the radiation energy density.
The temperature of the Universe begins to fall more slowly, $T \propto R^{-3/8}$, than
the usual $T \propto R^{-1}$.
 The Universe is never reheated, since the
temperature decreases at all times.
From $t \simeq t^{=}_\phi \, (\tau_\phi / t^{=}_\phi)^{3/5}$
till $t \simeq \tau_\phi$,
the entropy per comoving volume $S$ is growing $\propto R^{15/8}$.
At $\tau_\phi$ the Universe becomes purely radiation-dominated again with 
$T \propto R^{-1}$ and
a temperature
$T^\text{dec}_\phi =\left.T(\tau_\phi)\right|_{\text{rad-dom}}$, where we use
 the time--temperature relation for a radiation-dominated Universe,
\begin{equation}
\label{eq:tTinraddom}
 \left.t\right|_{\text{rad-dom}} = \left(\frac{45}{2 \pi^2 g_\ast (T)}\right)^{\frac{1}{2}} M_\text{Pl} T^{-2}  \, .
\end{equation}
This is the temperature after significant entropy production $T^{\text{after}}$,
which would be identified as the reheating temperature in the approximation of
simultaneous decay of all $\phi$ particles. We identify $T^\text{after} = T^\text{dec}_\phi$.
If $\rho_\phi$ never dominates over $\rho_\text{rad}$, $\phi$ decays never produce
a significant amount of entropy relative to the initial entropy.
Then the produced entropy is negligible.  However, if $\rho_\phi$
dominates over $\rho_\text{rad}$ before $\tau_\phi$, the produced
entropy dilutes significantly any relic density by a factor $\D$.

The dilution factor $\D$ is defined as the ratio of entropy per comoving volume 
after $\phi$ decay $S_f$ over the initial
 entropy per comoving volume $S_i$ and can be expressed as
\begin{equation}
\label{eq:d1}
 \D = \frac{S_f}{S_i} \simeq 0.82 \,
 \big\langle g_\ast^{1/3} \big\rangle^{3/4} \,
 \frac{m_\phi Y_\phi \tau_\phi^{1/2}}{M_\text{Pl}^{1/2}} \, ,
\end{equation}
where the angle brackets indicate the appropriately-averaged value of $g_\ast^{1/3}$ over the decay interval.
We see how $\D$ is determined by the properties of the unstable particle, i.e.\ its mass $m_\phi$ and 
lifetime $\tau_\phi$. Meanwhile it is assumed that $\tau_\phi > t^{=}_\phi$. The pre-decay abundance $Y_\phi$ of the unstable particle
depends on both its interactions and the earlier cosmology.

For convenience we would like to rephrase~\eqref{eq:d1} in terms of temperatures. Without entropy production after the
generation of the pre-decay abundance, it is constant till the particle decays. With $\rho= m n$ we find
\begin{equation}
\label{eq:help1}
 m_\phi Y_\phi = \frac{\rho_\phi}{s} = \left.\frac{\rho_\text{rad}}{s}\right|_{T=T^{=}_\phi}
 =\frac{3}{4} T^{=}_\phi \, ,
\end{equation}
where we have used~\eqref{eq:s}, \eqref{eq:rhorad} and 
$\rho_\phi = \rho_\text{rad}$ at $T^{=}_\phi$.

Using~\eqref{eq:tTinraddom} we can replace the particle lifetime in~\eqref{eq:d1} as
\begin{equation}
\label{eq:help2}
 \tau_\phi^\frac{1}{2} = \left(\frac{45}{2\pi^2}\right)^{\!\frac{1}{4}} g_\ast^{-\frac{1}{4}}(T^\text{dec}_\phi) \, M_\text{Pl}^\frac{1}{2} \, \bigl(T^\text{dec}_\phi\bigr)^{-1} \,.
\end{equation}
Plugging~\eqref{eq:help1} and~\eqref{eq:help2} into~\eqref{eq:d1} we obtain
\begin{equation}
\label{eq:d2}
 \D = 0.75 \,
 \frac{\big\langle g_\ast^{1/3} \big\rangle^{3/4}}{g_\ast^{1/4}(T^\text{dec}_\phi)}
 \, \frac{T^=_\phi}{T^\text{dec}_\phi} \,.
\end{equation}
This linear growth in temperature can also be expressed in terms of energy densities, since
\begin{equation}
 \rho_\phi = n_\phi m_\phi = s Y_\phi m_\phi = \frac{2\pi^2}{45} \, \frac{3}{4} \, g_\ast(T) \, T^3 \, T^{=}_\phi \,,
\end{equation}
where we have used~\eqref{eq:s} and~\eqref{eq:help1}. Taking this 
together with~\eqref{eq:rhorad} we find
\begin{equation}
  \frac{\rho_\phi}{\rho_\text{rad}} = \frac{T^{=}_\phi}{T} \, .
\end{equation}
Thus, for $T=T^\text{dec}_\phi$ we see that
\begin{equation}
 \frac{T^{=}_\phi}{T^\text{dec}_\phi} = \frac{\rho_\phi}{\rho_\text{rad}}(T^\text{dec}_\phi) \, ,
\end{equation}
where $\rho_\text{rad}$ is the density of the ``old'' radiation, i.e.\ it
does not include the radiation from $\phi$ decays.

The standard Big Bang model has been tested thoroughly up to
temperatures around $1\MeV$, where BBN occurs. 
Investigations of the thermalization of neutrinos
produced in $\phi$ decays or subsequent thermalization processes lead to lower limits on the temperature of 
the Universe after the entropy production $T^{\text{after}}$~\cite{Kawasaki:1999na}.
Neutrinos, which can thermalize through weak interactions only, are most important.
All other SM particles thermalize much faster due to their stronger interactions.  
The bounds found are in the range
\begin{equation}
\label{eq:Tafter}
 T^\text{dec}_\phi > T^\text{after}_\text{min} \simeq (0.7 \,\dots\, 4) \MeV \, 
\end{equation}
where weaker bounds come from BBN calculations~\cite{Kawasaki:2000en,Ichikawa:2005vw} and stronger bounds rely on the
neutrino energy density~\cite{Adhya:2003tr,Hannestad:2004px} 
exploiting overall best fits for cosmological parameters.
We take $T^\text{dec}_\phi \geq 4 \MeV \sim T_\text{BBN}$ as lower bound. 

Going back in time, thus towards higher temperatures, the first cosmological
event important for our scenario is the freeze-out of the NLSP\@.\footnote{
The QCD phase transition occuring between BBN and NLSP freeze-out seems not to
deliver any constraint on our scenario.}
Standard computations of relic abundances rely on the assumption of radiation domination during freeze-out.
If the Universe is dominated by matter during NLSP freeze-out,
the NLSP relic abundance is increased. Taking the later dilution by entropy production into account, 
the overall effect remains a reduction~\cite{Kamionkowski:1990ni}. 
The effects of different cosmological scenarios
on relic densities have been studied~\cite{Pallis:2004yy,Arbey:2008kv,Arbey:2009gt}
and there are computer codes~\cite{Arbey:2009gu}.  
In particular, the neutralino has been investigated,
also considering the production of neutralinos
in the decay of a dominating matter particle~\cite{Khalil:2002mu,Gelmini:2006pw,Gelmini:2006pq,Endo:2006ix}.
Since it is the easiest case to study, we take $T^{=}_\phi < T^\text{fo}_\text{nlsp}$.
Thereby the Universe is radiation-dominated during NLSP freeze-out
happening at $T^\text{fo}_\text{nlsp}$ and the
standard computations hold.%
\footnote{Using the simple estimate
 $H(T^\text{fo}_\text{nlsp}) \sim \Gamma(T^\text{fo}_\text{nlsp})$, where $\Gamma$ is the
 rate of NLSP annihilations, one finds that for
 $T^{=}_\phi = T^\text{fo}_\text{nlsp}$ the NLSP abundance is increased by a
 factor of only $\sqrt{2}$ compared to the standard case of radiation
 domination, while the freeze-out temperature stays nearly constant.
}
Later we will find that the window between BBN and NLSP freeze-out is
favoured intrinsically by the scenario.

Sticking to this particular window, we can evaluate~\eqref{eq:d2},
\begin{equation}
\label{eq:d3}
 \D \simeq 0.75 \times 10^3 \left(\frac{m_\text{nlsp}}{100 \GeV}\right) \left(\frac{4 \MeV}{T^\text{dec}_\phi}\right) ,
\end{equation}
 where we have plugged in $T^{=}_\phi=T^\text{fo}_\text{nlsp} \simeq m_\text{nlsp}/25$ and
 $\big\langle g_\ast^{1/3} \big\rangle \simeq 2.2 \simeq g_\ast^{1/3}(T^\text{dec}_\phi)$
with $g_\ast(T_\phi^\text{dec}) = 10.75$,
exploiting the fact that for $4 \MeV \leq T \leq 4 \GeV$ the effective relativistic degrees of freedom are known~\cite{Amsler:2008zzb}.
If we compare~\eqref{eq:d3} and~\eqref{eq:dboundfromTL}, we see that the cosmological window
between BBN and NLSP freeze-out is not only
the first and easiest but also sufficiently large to produce enough
entropy to come close to the upper
limit on $\D$ set by thermal leptogenesis itself.

This discussion assumes that there is no further entropy production after the generation of $Y_\phi$.
Otherwise, $Y_\phi$ would be diluted like any other relic abundance, i.e.\ $Y_\phi \to Y_\phi^\prime =\D_1^{-1} Y_\phi$.
There are two possibilities for the impact of such an earlier entropy increase $\D_1 > 1$. 
i) Despite the dilution, $\phi$ dominates the Universe for some time.
Then the later entropy production by the decay of $\phi$ is simply
reduced by a factor $\D_1$, as we see from~\eqref{eq:d1} since lifetime
and mass of the unstable particle are unchanged.
ii) The relic abundance of $\phi$ becomes so small that the particle
never dominates the energy density of the Universe. Then~\eqref{eq:d1}
does not hold and $S \simeq \text{const.}$, i.e.\ $\Delta = 1$.

After an arbitrary number of late events of entropy production $\D_i$ 
labeled by $i= 1,2,\ldots$, where the index 
implies a time-ordering with larger $i$ corresponding to later decays,
the total dilution factor is
\begin{equation}
\label{eq:dtot2}
 \D_\text{tot} = \prod_i \D_i \, .
\end{equation}
Here, ``late'' indicates that all decays happen after the freeze-out of all
unstable particles supposed to produce significant entropy, so that
their relic abundances are diluted by each earlier decay.
This implies
\begin{equation}
\label{eq:dk}
 \Delta_i = \max\left\{ \frac{\Delta_i(\Delta_{j<i}=1)}{\prod_{j<i} \Delta_j} \,, 1
 \right\} ,
\end{equation}
where $\Delta_i(\Delta_{j<i}=1)$ refers to the dilution factor obtained
from \eqref{eq:d1} without considering the other dilutions in the
calculation of $Y_\phi$.
As mentioned, we set $\D_i=1$, if a decaying particle does not come to
dominate the energy density of the Universe.  One can easily convince
oneself that the total dilution is simply given by the largest
individual dilution factor,
\begin{equation}
\label{eq:dtot1}
 \D_\text{tot} = \max \left\{ \Delta_i(\Delta_{j<i}=1) \right\} .
\end{equation}
The upper bound~\eqref{eq:dboundfromTL} limits $\D_\text{tot}$. 
The dilution of the NLSP abundance can be smaller than $\Delta_\text{tot}$ if
some decays happen before NLSP freeze-out.
Thus, we see from~\eqref{eq:dk} with~\eqref{eq:dboundfromTL} how our requirement of 
sufficient entropy production after NLSP freeze-out restricts the possibility of earlier entropy production.

\section{BBN Constraints on a Diluted Neutralino NLSP}
\label{sec:nlsp}
In this section we present constraints from Big Bang Nucleosynthesis on a neutralino NLSP in the case of
a gravitino with a mass of $m_{3/2}=100 \GeV$ being the LSP\@. We investigated those bounds
excluding the possibility of entropy production in~\cite{Covi:2009bk} and found for
masses below a~TeV a maximal gravitino mass of a few GeV\@. In the
following we assume that the neutralino is diluted after its freeze-out
by a factor $\D=10^3$.  It is trivial to infer the impact of arbitrary $\D$s.
BBN constraints on a stau NLSP with $\D$ up to $2 \times 10^4$ 
have been studied in~\cite{Buchmuller:2006tt,Pradler:2006hh}, where it
has been found that
interesting parameter regions are allowed for dilution factors $\D \sim 10^3$.

\begin{figure}[tb]
 \centering
   \includegraphics[width=0.49\linewidth]{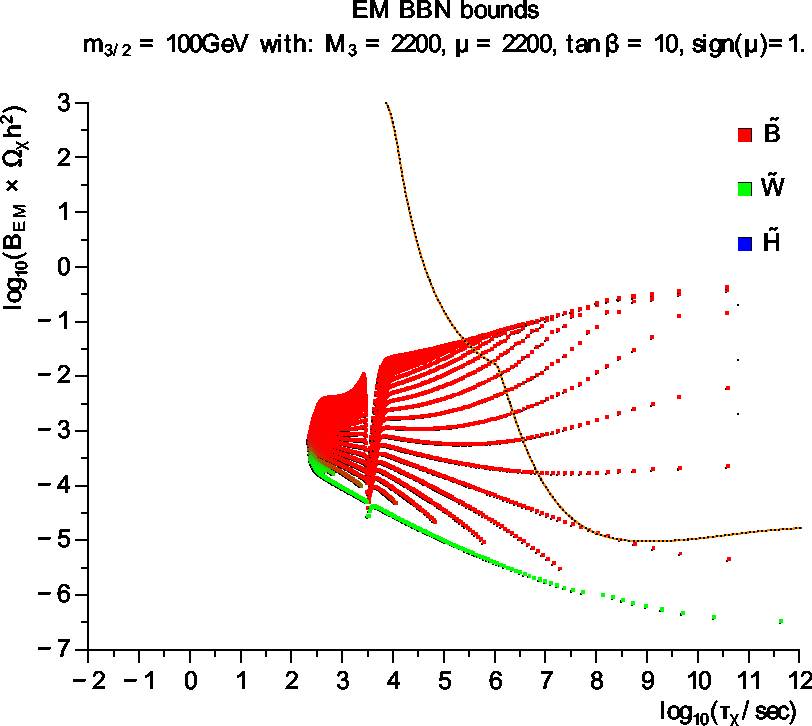}
   \hfill
   \includegraphics[width=0.49\linewidth]{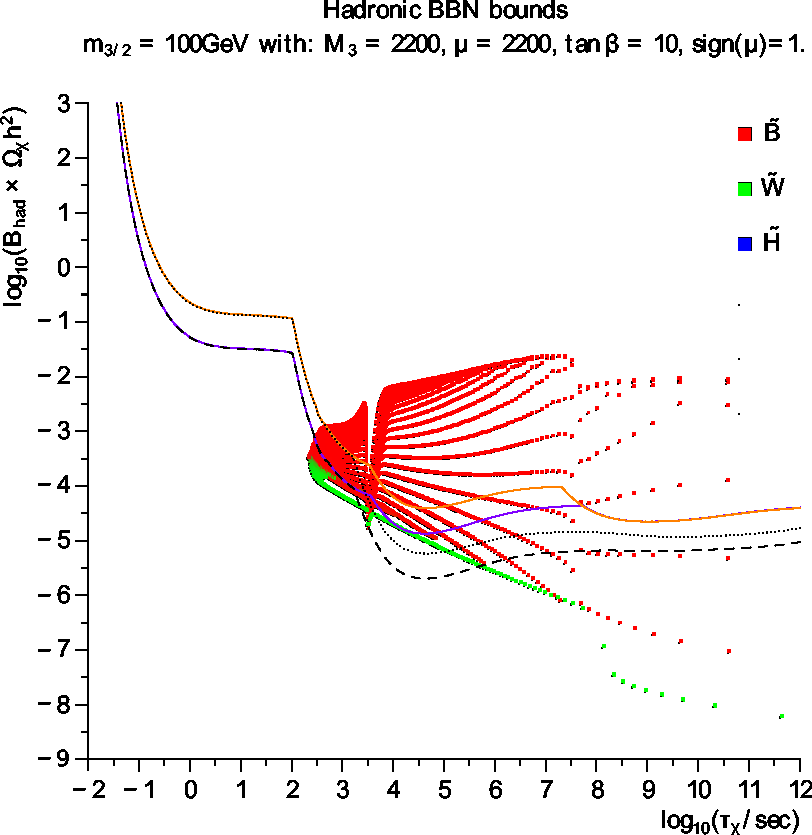}
\caption[BBN bounds on bino-wino]{\small Energy density of the bino-wino neutralino 
decaying into electromagnetic/hadronic products compared
  with the BBN electromagnetic (left) and hadronic constraints (right)
  for the case of a $100\GeV$ gravitino mass and a dilution factor $\D=10^3$.  The
  bounds are taken from~\cite{Jedamzik:2006xz}: the continuous (dashed) lines
  correspond to more (less) conservative bounds for the $^6$Li to
  $^7$Li ratio, and the region between the curves should not be
  considered as strictly excluded.  The red/upper and violet/lower
  curves in the hadronic plots are the constraints for $1\TeV$ and
  $100\GeV$ decaying particle mass, respectively.
  The mass increases from right to left as heavier particles decay
  faster.
  The composition goes from bino at the top to wino at the
  bottom while the colours give the dominant component.
  The deformation between the left and right
  panel is due to the mass dependence of the hadronic branching ratio
  with lighter NLSPs having lower branching ratios to hadrons. In
  contrast the electromagnetic branching ratio is always nearly one.}
\label{fig:bino-wino}
 \end{figure}

\begin{figure}[tb]
 \centering
   \includegraphics[width=0.49\linewidth]{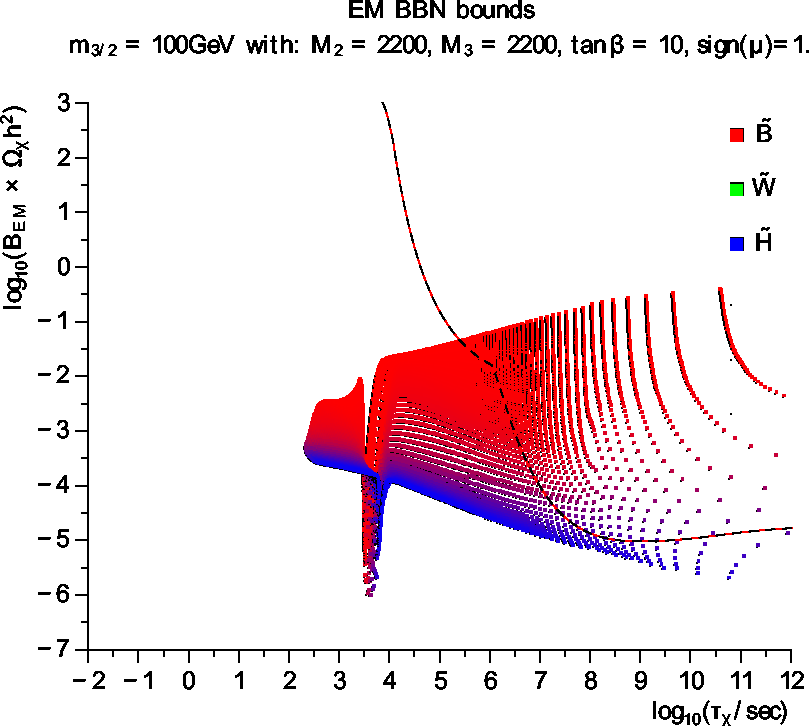}
   \hfill
   \includegraphics[width=0.49\linewidth]{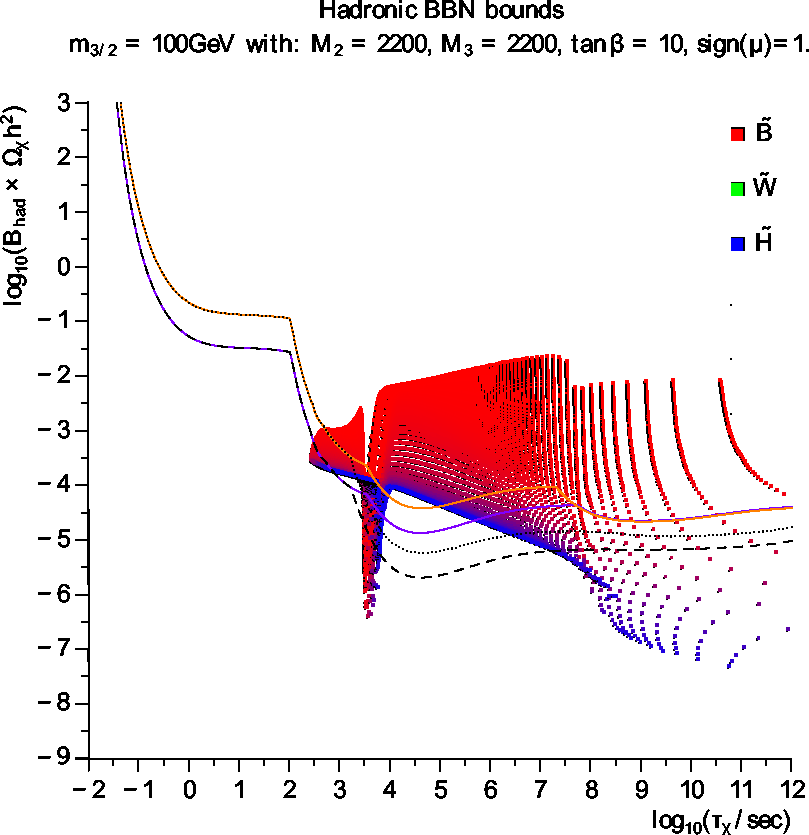}
\caption[BBN bounds on bino-Higgsino]{\small Energy density of the bino-Higgsino neutralino 
decaying into electromagnetic/hadronic products compared
  with the BBN electromagnetic (left) and hadronic constraints (right)
  for the case of a $100\GeV$ gravitino mass and a dilution factor $\D=10^3$.  The
  bounds are taken from~\cite{Jedamzik:2006xz}: the continuous (dashed) lines
  correspond to more (less) conservative bounds for the $^6$Li to
  $^7$Li ratio, and the region between the curves should not be
  considered as strictly excluded.  The red/upper and violet/lower
  curves in the hadronic plots are the constraints for $1\TeV$ and
  $100\GeV$ decaying particle mass, respectively. 
  The mass increases from right to left as heavier particles decay
  faster.
  The composition goes from bino at the top to Higgsino at the
  bottom while the colours give the dominant component.
  The deformation between the left and right
  panel is due to the mass dependence of the hadronic branching ratio
  with lighter NLSPs having lower branching ratios to hadrons. In
  contrast the electromagnetic branching ratio is always nearly one.}
\label{fig:bino-Higgsino}
 \end{figure}

\begin{figure}[tb]
 \centering
   \includegraphics[width=0.49\linewidth]{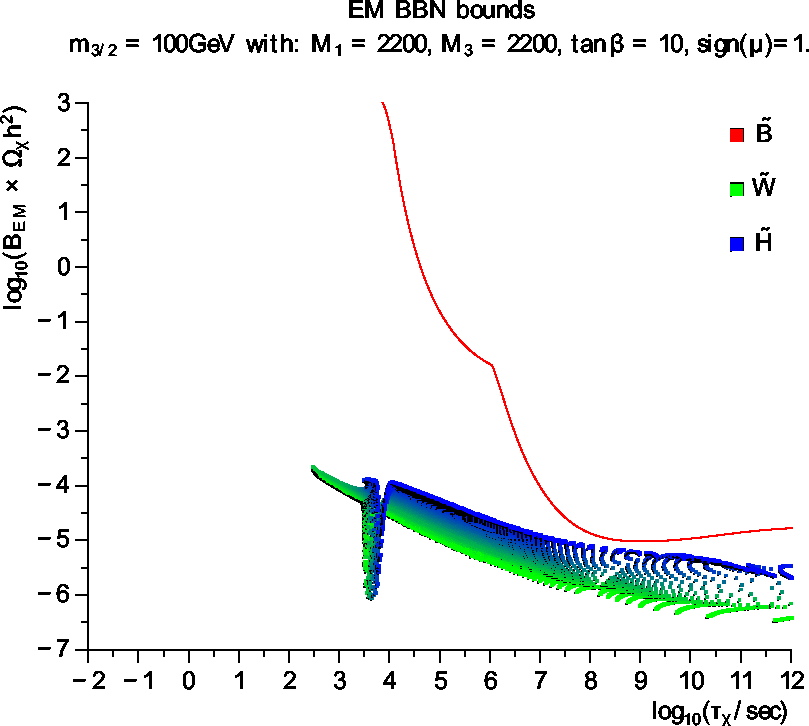}
   \hfill
   \includegraphics[width=0.49\linewidth]{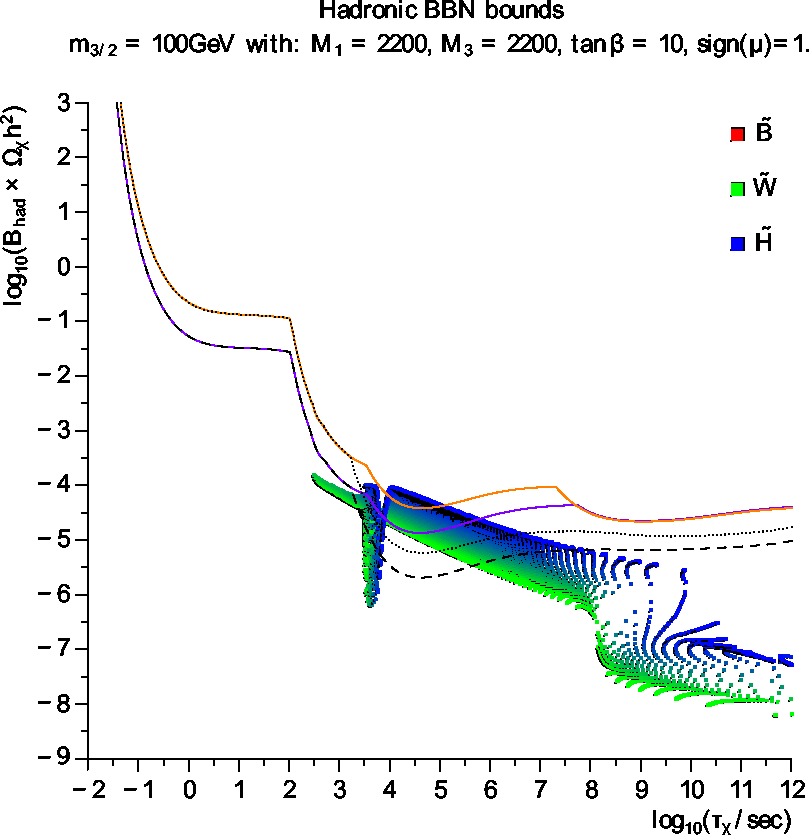}
\caption[BBN bounds on wino-Higgsino]{\small Energy density of the wino-Higgsino neutralino 
decaying into electromagnetic/hadronic products compared
  with the BBN electromagnetic (left) and hadronic constraints (right)
  for the case of a $100\GeV$ gravitino mass and a dilution factor $\D=10^3$.  The
  bounds are taken from~\cite{Jedamzik:2006xz}: the continuous (dashed) lines
  correspond to more (less) conservative bounds for the $^6$Li to
  $^7$Li ratio, and the region between the curves should not be
  considered as strictly excluded.  The red/upper and violet/lower
  curves in the hadronic plots are the constraints for $1\TeV$ and
  $100\GeV$ decaying particle mass, respectively.
  The mass increases from right to left as heavier particles decay
  faster.
  The composition goes from Higgsino at the top to wino at the
  bottom while the colours give the dominant component.
  The deformation between the left and right
  panel is due to the mass dependence of the hadronic branching ratio
  with lighter NLSPs having lower branching ratios to hadrons. In
  contrast the electromagnetic branching ratio is always nearly one.}
\label{fig:wino-Higgsino}
 \end{figure}

For a discussion of the neutralino decay channels, branching ratios and more details 
 we refer to~\cite{Covi:2009bk}.
To determine model-independent constraints within the MSSM
 we take all points that are not ruled out by LEP up to
a mass of 2~TeV, while we fix the masses of the sfermions to be above 2~TeV\@. 
To keep our analysis as general as possible
we do not fix all supersymmetric parameters according to
a specific scenario, but instead we set the soft SUSY breaking
parameters at the low energy scale. We keep the majority of
the parameters fixed and vary the gaugino and Higgsino mass parameters
to study how the lifetime and number density vary with the mass
and composition of the lightest neutralino.
We plot these points against the hadronic and electromagnetic BBN bounds
in Figs.~\ref{fig:bino-wino}--\ref{fig:wino-Higgsino}. The bounds are taken from~\cite{Jedamzik:2006xz}
and the different curves are explained in the figure caption.
The vertical axis corresponds to the fraction of the energy density that
decays to electromagnetic or hadronic products. A $\D >1$ shifts all points
downwards on this axis by a factor of $\D$. Therefore it is easy to infer
constraints for arbitrary $\D$ once a plot with fixed $\D$ is given.

Hadronic bounds are generally more constraining.
However, it has been found that large gravitino masses, for which light
neutralinos have a low hadronic branching ratio, are excluded by
the electromagnetic bounds.

In Fig.~\ref{fig:bino-wino} we consider a mixed bino-wino NLSP\@.
The large dip corresponds to resonant annihilation into the
pseudo-scalar Higgs, which happens for our choice of parameters 
at a neutralino mass $m_\chi \sim 1150 \GeV$. 
To increase $\eta_\text{B}^\text{max}$ we are more interested in the region of small
NLSP masses, since small NLSP masses allow more easily
 for small gluino masses in~\eqref{eq:etab2}.

Thanks to the dilution by entropy production the wino overcomes
the electromagnetic bounds for any mass even for masses close to the gravitino mass.
If the neutralino is mainly bino the electromagnetic bounds are more involved.
For a bino-like neutralino, masses below about $450 \GeV$ are excluded.
Smaller and smaller masses become allowed when the wino component
increases, so that there is allowed space for binos with a non-negligible
wino component and $m_\chi \sim 200 \GeV$ or even smaller masses.

The hadronic bounds exclude most of the parameter
space for a bino-wino with dominant bino component even with $\D=10^3$.
The mixed bino-wino states with $m_\chi \sim 200 \GeV$ mentioned above are found
on the less conservative ${}^6\text{Li}/{}^7\text{Li}$
exclusion line for a decaying particle of $100 \GeV$ mass. Thus we
find many points that should not be considered as strictly excluded
with masses around $200 \GeV$ and also mixed bino-wino states
that are allowed with masses smaller than $200 \GeV$.
 Winos with $m_\chi \lesssim 400 \GeV$ overcome even any
less conservative bound. For $400 \GeV \lesssim m_\chi \lesssim 1100 \GeV$
 the wino could violate the less conservative bound, 
 while even larger masses become allowed again.

Altogether, the situation is qualitatively different for bino and wino. While
the wino safely overcomes all bounds, especially at low masses, a bino-like
 neutralino with reasonable mass stays excluded even for much larger
dilution factors that would be in contradiction with successful thermal 
 leptogenesis~\eqref{eq:dboundfromTL}. However, there is also some space for bino-wino mixed
states that are mainly bino with masses below $200 \GeV$.
 
In Fig.~\ref{fig:bino-Higgsino} we consider a mixed bino-Higgsino NLSP\@.
The dip is broader in this case, since the Higgsino
component that couples to the pseudo-scalar Higgs is larger.
 Thanks to the dilution the Higgsino overcomes the electromagnetic bounds
like the wino for all masses, even though $\D$ should not be much
smaller than roughly $10^2$
 to allow for light Higgsino neutralinos. For the bino the situation is
comparable to the case of mixed bino-wino. No mixed bino-Higgsino state 
with a dominant bino component is allowed with masses as low as $200 \GeV$,
though.
 
 Again, the hadronic bounds exclude most of the bino parameter space. Exceptions
 are found in the dip and at very large masses. There are states with comparable
 bino and Higgsino components and $m_\chi \gtrsim 200 \GeV$---thus not excluded by the electromagnetic 
bounds---violating the less conservative hadronic bound. Higgsino neutralinos
lighter than $250 \GeV$ escape even these constraints, while they are excluded for
 $  670 \GeV \lesssim m_\chi \lesssim 1100 \GeV$.
 
 Altogether, we find that for mixed bino-Higgsino only states that are mainly
 Higgsino allow for preferrable small masses but then even down to the gravitino mass.
Considering only the conservative hadronic bound from the
${}^6\text{Li}$ to ${}^7\text{Li}$ ratio, in addition maximally mixed
states with masses in the region around $230 \GeV$ become allowed.
 $\D > 10^3$ would allow for larger bino components in the mixed bino-Higgsino.  
 
In Fig.~\ref{fig:wino-Higgsino} we consider a mixed wino-Higgsino NLSP\@.
Thanks to the dilution it overcomes the electromagnetic bounds for all masses and mixings.
 We are especially interested in the small mass region. By vertical shifts
 of all points
 we can search for the minimal dilution factor $\D^\text{min}$ to overcome the electromagnetic
bounds at small masses. A wino close to the gravitino mass 
becomes allowed for
 \begin{equation}
\D_{\widetilde W}^\text{min} \simeq 25  \, .
 \end{equation}
Larger $\D$ allows for more wino masses and eventually for light
Higgsinos at 
 \begin{equation}
\D_{\widetilde H}^\text{min} \simeq 90 \,.
 \end{equation}
 
Considering the hadronic constraints, winos with $m_\chi < 400 \GeV$ and
Higgsinos with $m_\chi < 200 \GeV$ satisfy all bounds. Wino-Higgsinos
with larger mass can be in conflict with the less
conservative bound, and the wino overcomes the more conservative one completely.
 Disregarding the dip, Higgsinos are excluded for a window $700 \GeV \lesssim m_\chi \lesssim 1300 \GeV$.
 
 In summary, entropy production after NLSP freeze-out can allow
 for a gravitino LSP of $100\GeV$ mass with a light neutralino NLSP\@.
 This reconciles thermal leptogenesis and gravitino dark matter within
 the scenario of a light neutralino NLSP\@. However, this depends
 on the composition of the lightest neutralino. The wino is
 the best case due to its small freeze-out abundance. Also a light
 Higgsino becomes allowed for reasonable dilution factors.
A light bino-like neutralino, which is typical for the Constrained MSSM, stays
 excluded for $\mgrav = 100 \GeV$ even if the possibility
 of entropy production after freeze-out is exploited.
 
Coming back to thermal leptogenesis, strictly speaking none of the
points in the figures allows for a sufficiently high reheating
temperature, since the gluino mass has been fixed at $2.2\TeV$.
However, in the parameter space regions with smaller neutralino masses,
the gluino mass can be lowered without affecting our considerations at
all.  Consequently, all allowed points with a neutralino mass below a
TeV can be compatible with thermal leptogenesis.

\section{Search for a viable Candidate}
\label{sec:candidate}
\subsection{General Requirements}
In this section we discuss candidates for the
entropy-producing particle $\phi$ of the previous sections. 
After enumerating the required properties in general,
we exemplify in detail an implementation of the scenario with the axion
multiplet.

To dilute the NLSP relic density, $\phi$ must \emph{i)~decay after
NLSP freeze-out.}  But, for sure, it \emph{ii)~decays before BBN\@.}
Thus the lifetime $\tau_\phi$ or equivalently the
decay temperature $T^\text{dec}_\phi$ is constrained to a window.
The particle has to be \emph{iii)~produced in the early
Universe such that it dominates the energy density before
BBN\@.}  Meanwhile we stick to the case where its relic
density \emph{iv)~does not come to dominate before NLSP freeze-out.}
Thus the relic density prior to its decay $\rho_\phi = Y_\phi m_\phi s$ is also constrained to a window.
Requirements iii)+iv) imply that the dominance of $\phi$ has to grow
with the expansion, which is true for non-relativistic matter.
So $\phi$ is implicitly assumed to become non-relativistic before BBN\@.
The requirements i)+ii) and iii)+iv) constrain two different
quantities $\tau_\phi$ and $\rho_\phi$, which are determined in different ways
but by the same properties of $\phi$, namely its couplings and mass.

Requirements i)+ii) constrain only the total decay rate $\G_\phi^\text{tot}=\tau_\phi^{-1}$.
In fact, the branching ratios of $\phi$ into the LSP and NLSP are also
constrained. The branching ratio into the NLSP $B_{\phi \to \text{nlsp} + \ldots}$
must be so small that the \emph{v)~NLSP decay problem is not
reintroduced by the decay.}
Branching ratios into the LSP are always restricted by overproduction.
Especially when the LSP is already produced thermally as in our
scenario of thermal leptogenesis with gravitino dark matter, $\phi$ should
\emph{vi)~not produce too many LSPs in its decays.}
Since $\phi$ even dominates the energy density of the Universe at its
decay, the requirements v)+vi) force the corresponding branching ratios
to be---at least---close to zero.

In addition, $\phi$ must be \emph{vii)~compatible with gravitino dark
matter.}  For example, the gravitino would become unstable due to the
existence of $\phi$, if it could decay into $\phi$.  This would take
away the explanation for the observed dark matter abundance, if the
gravitino lifetime were too short, or by itself be in conflict with
other observations.

Finally, \emph{viii)~unavoidable by-products of $\phi$ have to be harmless.}
For example, such by-products are
the supermultiplet partners in SUSY\@. They are harmless, if they do not
violate ii) or vii), are free of the problems
solved by v)+vi) and do not introduce new
problems on their own.

\begin{table}
  \label{tab:req}
\renewcommand{\arraystretch}{1.3}
\begin{center}
\begin{tabular}{ccc}
\hline\hline
No.       	&	 Requirement 				& 		Comment
\\ \hline
i		& $T^\text{dec}_\phi < T^\text{fo}_{\text{nlsp}}$		& to have effect on $\O_{\text{nlsp}}$
\\
ii		& $T^\text{dec}_\phi > T_{\text{BBN}}$		& not to spoil BBN
\\[1mm]
iii		& $\frac{\rho_\phi}{\rho_\text{rad}}(T^\text{dec}_\phi) > 1$ & $\mathcal{O}(10) < \D < 10^4$
\\
iv		& $\frac{\rho_\phi}{\rho_\text{rad}}(T^\text{fo}_\text{nlsp}) < 1$ & for standard NLSP freeze-out
\\[1mm]
v		& $B_{\phi \to \text{nlsp} + \ldots} \simeq 0$ & from NLSP decay problem 
\\
vi		& $B_{\phi \to \gr + \ldots} \simeq 0$ 	& from overproduction ($\O_{3/2}^\text{tp} \simeq \O_\text{DM}$)
\\[1mm]
vii		& e.g.\ $\tau_{3/2} \gg t_0$			& compatibility with gravitino dark matter 
\\[1mm]
viii		& ii) and v)--vii) 				& for by-products; no new problems
\\
\hline\hline
\end{tabular}
\end{center}
\caption{\small List of requirements for our scenario of entropy produced by $\phi$ to dilute the NLSP.}
\end{table}
Altogether, the properties of $\phi$ seem to be highly
constrained. We summarise the requirements in Tab.~\ref{tab:req}.
The number of free parameters---mass and couplings---is finite.
Since they enter in different ways for different constrained quantities,
 it is not a matter of course that the scenario of late-time
 entropy production is viable at all. Especially if $Y_\phi$ is
 produced
 thermally via scatterings, the  same coupling might be responsible
 for the production and the late decay.

On the other hand, many extensions of the SM contain or predict
super-weakly interacting and hence long-lived particles.
Such particles generically satisfy i), if not by definition.
In order to ensure that they are harmless, one usually 
demands that they decay before BBN conform to ii). 
Thermal leptogenesis places the upper limit~\eqref{eq:dboundfromTL} on
the maximally allowed dilution, which implies that for decay right
before BBN iv) has to hold at least approximately.
Considering high reheating temperatures and the growth of
$\rho_\text{mat}/\rho_\text{rad} \propto R$,
it is probable that the energy density of late-decaying particles 
dominates over the radiation energy density at their decay.
Thus, iii) can be considered as fulfilled generically, which in fact
normally poses a problem. 
Besides, the decay into superparticles usually has to be suppressed
in order to avoid producing too much dark matter and further
late-decaying particles like the NLSP or the gravitino, in case it is
not the LSP\@.  Consequently, v) and vi) are generic, too, possibly
amended by $B_{\phi \to \text{lsp} + \ldots} \simeq 0$, if the gravitino
is not the LSP\@.
In any case the scenario has to be compatible with whatever
is supposed to form the dark matter, so that vii) is generic.
Also viii) arises as a generic requirement on any
late-decaying particle and is particularly constraining in
supersymmetric models.

In summary, $\phi$ is severely constrained such that the scenario of
entropy production to dilute the NLSP density might appear unappealing.
However, in extensions of the SM containing long-lived particles, in
principle i)+ii) and v)--viii) are no new requirements and are present---in
appropriate form---without considering entropy production at all.  If
$T^\text{dec}_\phi \sim T_\text{BBN}$, successful thermal leptogenesis
favours the situation of iv).  Finally, for the corresponding high
reheating temperatures iii) is generic.
Thus, all the requirements of Tab.~\ref{tab:req} either have to be
fulfilled or are generically fulfilled.  In other words, the solution of
the generic problems of long-lived particles may well cause the entropy
production desired to solve the NLSP decay problem and thereby reconcile
thermal leptogenesis and gravitino dark matter.

With a specific candidate at hand the details have to
be worked out. One has to determine whether a candidate is excluded,
not useful or can be the solution and how generically this is true.
As an example, we investigate the axion multiplet, which
is motivated by a completely disconnected problem of the SM.

\subsection{Example: Axion Multiplet}
The strong CP problem of the SM can be solved by
the Peccei-Quinn (PQ) mechanism~\cite{Peccei:1977hh,Peccei:1977ur}. 
An additional global
U$(1)_\text{PQ}$ symmetry referred to as PQ symmetry, which is broken
spontaneously at some PQ scale, can explain the smallness
of the CP-violating $\Theta$-term in QCD\@.
The pseudo Nambu-Goldstone boson
associated with this spontaneous symmetry breaking is
called axion~\cite{Weinberg:1977ma,Wilczek:1977pj}. It has not yet been observed.
However, axion physics provides a lower limit~\cite{Amsler:2008zzb,Raffelt:2006cw}
on the axion decay constant,
\begin{equation}
\label{eq:fa}
 f_a \gtrsim 6 \times 10^8 \GeV \, .
\end{equation}
We identify the PQ scale with the axion decay constant.%
\footnote{This is equivalent to the choice $N=1$, where $N$ characterizes
the colour anomaly of U$(1)_\text{PQ}$.  With this choice we also avoid
possible problems with topological defects.
  In the working scenario at the end the axion abundance from strings
  is negligible.
  For $N \neq 1$ our formulae
hold, but $f_a = f_\text{PQ}/N \neq f_\text{PQ}$.
}
Since our considered reheating temperatures are
relatively large, it is probable that $T_\text{R}>T_\text{PQ}\sim f_a$.

If the PQ mechanism is supersymmetrised~\cite{Nilles:1981py,Kim:1983dt}, the
axion $a$ is part of a supermultiplet, the axion multiplet.
It consists of the axino $\widetilde a$ containing 
the fermionic degrees of freedom, the axion itself and
the saxino or saxion $\phi_\text{sax}$, which is an additional real scalar
degree of freedom.

We investigate $\phi_\text{sax}$ as candidate for the entropy-producing
particle $\phi$ of the previous sections. Thus
the axion and the axino are unavoidable by-products, while
it will become clear why they are no candidates themselves.
However, the saxion is motivated from the strong CP problem
and not from our scenario.

\subsubsection{Thermally Produced Multiplet}
After reheating, reactions 
like $q \bar q \leftrightarrow  g \phi_\text{sax}$
and $gg \leftrightarrow  g \phi_\text{sax}$ drive
the saxion into thermal equilibrium
if the reheating temperature is larger than its decoupling temperature~\cite{Kim:1992eu}
\begin{equation}
\label{eq:tsaxdcp}
 T_\text{sax}^\text{dcp} \simeq 10^{11} \GeV \left(\frac{f_a}{10^{12} \GeV}\right)^2 \left(\frac{0.1}{\alpha_\text{s}}\right)^3  \, ,
\end{equation}
where $\alpha_\text{s}=g_s^2(\mu) / 4\pi$ evaluated at the scale relevant for
the processes under consideration. The equilibrium saxion abundance is
given by
\begin{equation}
\label{eq:ysax}
 Y_\text{sax}^\text{eq}= \frac{45\,\zeta(3)}{2 \pi^4 g_\ast(T^\text{dcp}_\text{sax})}
 \simeq  1.21 \times 10^{-3} \, .
\end{equation}
Throughout this paper we assume for simplicity the particle content of the MSSM when 
we determine for example the relativistic degrees of freedom
in the Universe. The numerical changes from adding the axion multiplet,
for instance, are tiny as we can see in all estimates.

The saxion becomes non-relativistic at a temperature
$T^\text{nr}_\text{sax} \simeq 0.37 \, m_\text{sax}$ around its mass~\cite{Kim:1992eu}. 
From $0.37 \, m_\text{sax} \simeq T^\text{nr}_\text{sax} > T^{=}_\text{sax} = T^\text{fo}_\text{nlsp} \simeq m_\text{nlsp}/25$
would arise a lower bound
\begin{equation}
\label{eq:msaxminnr}
 m_\text{sax} > \frac{m_\text{nlsp}}{9.25} \, .
\end{equation}
We will find that this is weaker than the lower bound on the saxion mass from early
enough decay~\eqref{eq:msaxmin}
and thus in nearly all cases and at least in the interesting ones does not yield any constraint. 
With~\eqref{eq:help1} we see
that if the saxion lives long enough, it dominates the energy density of the Universe below the temperature
\begin{equation}
\label{eq:tsaxeq}
 T_\text{sax}^{=} = \frac{4}{3} Y_\text{sax}^\text{eq} m_\text{sax} \simeq  1.6 \GeV \left(\frac{m_\text{sax}}{1 \TeV} \right) \, .
\end{equation}
We avoid matter domination during NLSP freeze-out by requiring $T_\text{sax}^{=} < T_\text{nlsp}^\text{fo} \simeq m_\text{nlsp}/25$,
which gives an upper bound
\begin{equation}
\label{eq:msaxmax}
 m_\text{sax} <  2.5 \TeV \left(\frac{m_\text{nlsp}}{10^2 \GeV}\right)\, .
\end{equation}
As we know~\eqref{eq:d2}, considerable entropy is produced only, if
$T_\text{sax}^\text{dec} \ll T_\text{sax}^{=}$.  The saxion decay temperature can be 
derived from \eqref{eq:tTinraddom} with $T=T_\text{sax}^\text{dec}$ and
$t=1/\Gamma^{gg}_\text{sax}$, where 
\begin{equation}
\label{eq:widthsaxgg}
 \G^{gg}_\text{sax} \simeq \frac{\alpha_\text{s}^2 m_\text{sax}^3}{128 \pi^3 f_a^2}
\end{equation}
is the width of the dominant saxion decay into two
gluons~\cite{Kim:1992eu}.  This yields~\cite{Lyth:1993zw}
\begin{equation}
\label{eq:tsaxdec}
 T_\text{sax}^\text{dec} \simeq 53 \MeV \left(\frac{10^{12} \GeV}{f_a}\right) \left(\frac{m_\text{sax}}{1 \TeV}\right)^{\frac{3}{2}} 
 \left(\frac{\alpha_\text{s}}{0.1}\right) \left(\frac{10.75}{g_\ast(T^\text{dec}_\text{sax})}\right)^\frac{1}{4}  \, .
\end{equation}
Here, $\a_\text{s}$ has to be evaluated at $m_\text{sax}$. 
As we do not
consider an extremely large range of saxion masses,
$\a_\text{s}(m_\text{sax})$ does not vary significantly. 
Besides, in the range of parameters considered,
$g_\ast(T^\text{dec}_\text{sax})$ remains approximately constant.
Therefore, we drop the explicit dependence on $\a_\text{s}$
and $g_\ast(T^\text{dec}_\text{sax})$
in the following equations.
Together with the bound~\eqref{eq:Tafter} from early enough decay, we obtain the lower limit
\begin{equation}
\label{eq:msaxmin}
 m_\text{sax}>  180 \GeV \left(\frac{T^{\text{after}}_\text{min}}{4 \MeV}\right)^{\frac{2}{3}} \left(\frac{f_a}{10^{12} \GeV}\right)^{\frac{2}{3}} \, .
 % \left(\frac{0.1}{\alpha_\text{s}}\right)^\frac{2}{3} \left(\frac{g_\ast (T^\text{dec}_\text{sax})}{10.75}\right)^\frac{1}{6}
 % uncomment if functional dependence should be made visible. However the values do practically not vary.
\end{equation}
If we compare this lower bound with~\eqref{eq:msaxminnr}, we see that~\eqref{eq:msaxmin} is stronger as long as
\begin{equation}
\label{eq:msaxhelp}
\left(\frac{f_a}{1.5 \times 10^{10} \GeV}\right)^{\frac{2}{3}} \gtrsim \left(\frac{m_\text{nlsp}}{10^2 \GeV}\right)\, .
\end{equation} 
In any case, the saxion mass is constrained to a window. 
Since there is no additional source of
SUSY breaking, one expects $m_\text{sax} \sim m_\text{susy}$ and thus a saxion mass in the TeV range,
i.e.\ $10^2 \GeV \lesssim m_\text{sax} \lesssim 1 \TeV$. Thus, from this discussion
one might conclude that the requirements i)--iv) of Tab.~\ref{tab:req}
are naturally fulfilled.

Plugging~\eqref{eq:tsaxeq} and~\eqref{eq:tsaxdec} into~\eqref{eq:d2} we obtain 
\begin{equation}
\label{eq:d4}
 \D \simeq 13 \, \big\langle g_\ast^{1/3} \big\rangle^{3/4} 
 \left(\frac{f_a}{10^{12} \GeV}\right) \left(\frac{1 \TeV}{m_\text{sax}}\right)^\frac{1}{2}\, .
\end{equation}
For simplicity we replace $\big\langle g_\ast^{1/3} \big\rangle$
for the moment by $2.2$, the value estimated for~\eqref{eq:d3}.
 We plug in the bounds on the saxion mass~\eqref{eq:msaxmax} and~\eqref{eq:msaxmin} 
to find
\begin{equation}
\label{eq:drange}
 14 \left(\frac{f_a}{10^{12} \GeV}\right) \left(\frac{10^2 \GeV}{m_\text{nlsp}}\right)^\frac{1}{2}  <
 \D < 55 \left(\frac{f_a}{10^{12} \GeV}\right)^\frac{2}{3} \left(\frac{4 \MeV}{T^{\text{after}}_\text{min}}\right)^\frac{1}{3} \, .
\end{equation}
The lower bound on $\D$ shows that~\eqref{eq:msaxmin} is always stronger than~\eqref{eq:msaxminnr}, since
the inequality~\eqref{eq:msaxhelp} is always true as long as significant entropy is produced and if
this were not the case, \eqref{eq:msaxminnr} would not be considered at all.

If saxions are part of the particle spectrum,
 \eqref{eq:drange} 
shows two things.
i) It is likely that saxions produce significant entropy in their decays.
To avoid it, one would have to restrict the reheating temperature such that they never
enter equilibrium, or to choose safe values for $f_a$ and $m_\text{sax}$, e.g.\ $m_\text{sax}=1 \TeV$ 
and $f_a=10^{10} \GeV$. 
ii) The corresponding dilution factor is much smaller than the maximal value
allowed by cosmology and preferred as a solution of the NLSP decay problem.

The dilution factor can be increased by a larger axion decay constant,
which makes the saxion more weakly interacting. 
From~\eqref{eq:drange} we see that we need
$f_a \simeq 5.2 \times 10^{13} \GeV$ to reach the maximum
$\D \simeq 0.75 \times 10^3$ of~\eqref{eq:d3}. 
This increases the decoupling temperature~\eqref{eq:tsaxdcp} and thereby
the reheating temperature required to have the saxions in thermal equilibrium. 
If they did not enter equilibrium, the abundance would be $Y_\text{sax} \ll Y_\text{sax}^\text{eq}$,
and the saxion would be useless for our purpose.
From the requirement $T_{\text{R}} > T_\text{sax}^\text{dcp}$
and~\eqref{eq:tsaxdcp} we derive the upper bound
\begin{equation}
 f_a \lesssim 1.0 \times 10^{12} \GeV \left(\frac{T_{\text{R}} }{4 \times 10^{12} \GeV} \right)^\frac{1}{2}
   \left(\frac{\alpha_\text{s}}{0.03}\right)^\frac{3}{2} \, ,
\end{equation}
where $\alpha_\text{s} (4 \times 10^{12} \GeV) \simeq 0.03$. Already such a $T_{\text{R}}$
corresponds---at least in the case of heavy gravitinos---to an allowed but relatively
large dilution factor $ \Delta \sim 10^3$, cf.~\eqref{eq:o32est} and~\eqref{eq:dboundfromTL}. For the small $\D$s of~\eqref{eq:drange}
the situation becomes worse and is in fact inconsistent with itself.

To sum up, if thermally produced saxions are to deliver the desired
entropy, we need a large axion decay constant.  Then we also need a
large reheating temperature to make the saxion enter thermal equilibrium.
This results in an overproduction of gravitinos, if they are produced
without entering equilibrium, so that the scenario is not viable.

On the other hand, if the gravitino is so light that it enters equilibrium after
reheating, the relic gravitino density becomes independent of
the reheating temperature as mentioned in Sec.~\ref{sec:entropy}.
Moreover, there could be another saxion production mechanism, which is
the alternative we will concentrate on in the next section. 

Let us therefore continue discussing the requirements of
Tab.~\ref{tab:req}, turning to the decay products of the saxion.
Due to R-parity conservation, it
must produce sparticles in pairs and thus cannot decay into
single gravitinos. Besides, the decay into gravitino pairs is
negligible, since it is suppressed by an additional factor of $M_\text{Pl}^2$.
Consequently, requirement vi) is satisfied without any effort.

To fulfill requirement v) the decay into any other sparticle pair must be 
kinematically forbidden,
i.e.\ $m_\text{sax}< 2\,m_\text{nlsp}$.  This is the case if the saxion is lighter
or not much heavier than the gravitino.  Given that one expects both
the gravitino and the saxion mass to be of order $m_{susy}$, such a
spectrum does not seem unlikely.
It is understood that requirement v) does not apply for a light
gravitino produced in thermal equilibrium, since the NLSP
decays early enough before BBN and does not overproduce gravitinos.

On the other hand, the saxion may decay as well into
two axions with~\cite{Chun:1995hc}
\begin{equation}
\label{eq:saxtoaa}
 \G^{aa}_\text{sax} \simeq \frac{x^2 m_\text{sax}^3}{32 \pi f_a^2} \, ,
\end{equation}
where the self-coupling $x$ can be of order $1$. In that case the Universe
would be filled with relativistic axions during the process of
entropy production. This would change
the effective number of neutrino species, which
could spoil the success of BBN\@.
This requires $x \ll 1$ and there are concrete models realizing
this~\cite{Chun:1994zp}.

Up to now, we went through the requirements i)--vi) of Tab.~\ref{tab:req}.
We do not see any incompatibilities between the saxion producing 
entropy and gravitino dark matter. Hence, vii) is
fulfilled automatically as well.
Facing viii) we have to take care of the unavoidable by-products.

\paragraph{Axion}
Due to the similar coupling strength the axion also enters equilibrium, if
the saxion does. Then its thermal abundance is the same as the
saxion abundance~\eqref{eq:ysax}. Due to its tiny mass~\cite{Kolb:1990vq}
\begin{equation}
 m_a \simeq 0.62 \meV \left(\frac{10^{10}\GeV}{f_a}\right)
\end{equation}
its thermal relic density is negligibly small. 
The axion decays harmlessly into two photons with a lifetime
many orders of magnitude
 larger than the age of the Universe~\cite{Kolb:1990vq}, while 
 the gravitino cannot decay into axions due to R-parity.
 
Axions are also produced by vacuum misalignment, which leads to the
bound $f_a \lesssim 10^{12} \GeV$ to avoid
overproduction~\cite{Kolb:1990vq}.
However, for considerable entropy production by the saxion this bound
no longer holds, since the Universe is dominated by the saxion at the
onset of axion oscillations.  Then the axion density is given
by~\cite{Kawasaki:1995vt}
\begin{equation}
 \O_a h^2 \simeq 0.21 \left(\frac{T_\text{sax}^\text{dec}}{4 \MeV}\right) \left(\frac{f_a}{10^{15} \GeV }\right)^2 .
\end{equation}
If we require $\O_a/\O_\text{DM} = r \ll 1$, we find
\begin{equation}
\label{eq:oabound}
 \left(\frac{f_a}{10^{14} \GeV}\right)^2 \lesssim
 \left(\frac{r}{0.02}\right) \left(\frac{4 \MeV}{T_\text{sax}^\text{dec}}\right) ,
\end{equation}
so that values of $f_a > 10^{12} \GeV$ are indeed allowed. The bound from $\O_a \ll \O_{3/2} \simeq \O_{\text{DM}}$ is self-consistently
cured by the decaying saxion.

Altogether, there is no problem at all with the axion in our scenario.

\paragraph{Axino}
This is different for the axino $\widetilde a$.
Due to SUSY the axino has couplings similar to the 
saxion couplings and thus the same decoupling
temperature~\eqref{eq:tsaxdcp} as the saxion and
 a similar relic abundance.
The natural mass range for the axino is 
$\mathcal{O}(\text{keV})< m_{\widetilde a} < \mathcal{O}(\mgrav) \sim m_\text{susy}$~\cite{Chun:1992zk,Chun:1995hc}.
We demand
$m_{\widetilde a} > \mgrav$ to keep the gravitino as LSP\@. Then
the mass range for the axino becomes similar to that of the saxion.
In comparison to the saxion it has the opposite R-eigenvalue
and thus must produce sparticles in its decays.
With a light gravitino and
axino NLSP, one would arrive at another NLSP decay problem.
The situation would be worse than our starting point.
If the axino should not produce gravitinos, which would lead to
$Y_{3/2} \sim Y^\text{eq}_{\widetilde a}=Y^\text{eq}_{3/2}$, there
must be another decay
channel kinematically open, i.e.\ $m_{\widetilde a} > m_\text{nlsp}$. 
In the most 
interesting case $ m_\text{nlsp}$ is close to the gravitino mass and
the axino fulfills requirement vi). 

If kinematically allowed, the axino is expected to decay dominantly into a gluino-gluon pair with
\begin{equation}
 \G^{\widetilde g g}_{\widetilde a} \simeq
 \frac{\alpha_\text{s}^2 m_{\widetilde a}^3}{128 \pi^3 f_a^2}\, . 
\end{equation}
From the requirement of early enough decay we derive the same lower
bound on $m_{\widetilde a}$ as found for the saxion~\eqref{eq:msaxmin}.
This would allow an axino lighter than the expected gluino mass. Weaker decays into sparticles and SM particles
were to investigate. However, all these processes finally produce
the lightest ordinary supersymmetric particle.
The case of a heavy axino that decays after NLSP
freeze-out has been studied in~\cite{Choi:2008zq} for a
neutralino dark matter scenario including the weaker decay into a neutralino
and the re-annihilation of neutralinos.
Even if the axinos are not in thermal equilibrium after inflation,
they are regenerated by thermal scatterings and decays
in the thermal plasma. The density produced in this way 
can be estimated in units of today's critical density as~\cite{Covi:2001nw,Brandenburg:2004du,Strumia:2010aa}
\begin{equation}
\label{eq:oaxino}
 \O_{\widetilde a} h^2 \simeq 7.8 \times 10^2 \, \D^{-1}
 \left(\frac{m_{\widetilde a}}{1 \TeV}\right) \left(\frac{T_\text{R}}{10^9 \GeV}\right)   
 \left(\frac{10^{14} \GeV}{f_a}\right)^2 .
\end{equation}
The resulting neutralino density is many
magnitudes larger than the thermal 
relic abundance (cf.\ e.g.\ Fig.~\ref{fig:bino-wino}), which in our
scenario reintroduces the NLSP decay problem. 
In fact, the problem becomes much worse.
Thus requirement v) is badly violated by the axino.

If we require the axino to decay before NLSP freeze-out, so that
we do not have to care about the produced number of NLSPs since they
thermalize normally, 
we find by a derivation analogous to that of~\eqref{eq:msaxmin} the lower bound on the axino mass
\begin{equation}
\label{eq:maxinomin2}
 m_{\widetilde a} \gtrsim 1.2 \times 10^2 \TeV \left(\frac{m_\text{nlsp}}{10^2 \GeV}\right)^{\frac{2}{3}} 
 \left(\frac{f_a}{10^{13} \GeV}\right)^{\frac{2}{3}}
  \left(\frac{0.1}{\alpha_\text{s}}\right)^{\frac{2}{3}}
   \left(\frac{g_\ast(T^\text{dec}_{\widetilde a})}{100}\right)^{\frac{1}{6}} \, .
\end{equation}
Since the gravitino problem could also be solved by making the gravitino
comparably unnaturally heavy, such a large axino mass is not considered as 
a solution here. Furthermore, such an axino would produce considerable
entropy with $\D_{\widetilde a} \simeq 290$. 
From
the discussion at the end of Sec.~\ref{sec:EntropyFromDecays}
we know that this would spoil our scenario, since $\Delta_{\widetilde a}$ would dilute the saxion but not the NLSP\@. Thus, the situation
is also inconsistent. The required axino mass to achieve $\D_{\widetilde a}=1$
would be larger than about $10^7 \TeV$.

Altogether, requirement viii) of Tab.~\ref{tab:req} is
badly violated by the axino.  Consequently, the thermally
produced saxion---and obviously also the axino itself---is ruled out as
viable particle to produce significant entropy after
NLSP freeze-out.
The exception to this conclusion is
a light gravitino in thermal equilibrium after reheating,
since it allows for high reheating temperatures and the 
NLSP decay problem is absent.

One may worry then if the strong
CP problem can be solved by the Peccei-Quinn mechanism
in scenarios of standard thermal leptogenesis
with very light gravitino dark matter only. Going through the
equations, especially from~\eqref{eq:maxinomin2},
we see that the axino becomes harmless for smaller
axion decay constants $f_a \lesssim 10^{10} \GeV$
with an acceptable axino mass $m_{\widetilde a} \gtrsim 1.2 \TeV$.
Since its decay into the gravitino is suppressed
like $(f_a/M_\text{Pl})^2$, the contribution to the
gravitino density from axino decay is negligible.
However, by inspection of~\eqref{eq:d4} we
see that in this case the saxion is unable to produce a
significant amount of entropy.  Then also the axion abundance restricts
$f_a$ to values smaller than about~$10^{10} \GeV$. 

In summary,
 by making the axino harmless
we find that the thermally produced multiplet may also
  exist in scenarios of thermal
leptogenesis with gravitino dark matter
that does not enter equilibrium
after reheating. However, the axion decay constant
is restricted to a small window.
Moreover, the thermally produced multiplet is
in fact useless for our purpose.
This is due to two generic features of
the considered scenario: i) 
Superpartners have similar
couplings and masses. ii)~The same
coupling---or at least couplings of the
same strength---were responsible for
production and late
decay of the entropy-producing particle.

\subsubsection{Generic Thermally Produced Particle}

The negative result for the saxion can be generalized to other
late-decaying particles that are produced in thermal equlibrium by
processes controlled by the same coupling as the decay.
As the simplest estimate, let us assume that the particle $\phi$ under
consideration couples to SM particles via non-renormalisable
interactions suppressed by an energy scale $\Lambda$ and that the rate
of reactions keeping $\phi$ in thermal equilibrium at high temperatures
can be written as
\begin{equation}
 \Gamma_\phi^\text{prod} = x \, \frac{T^3}{\Lambda^2} \,,
\end{equation}
where $x$ is a model-dependent, dimensionless quantity containing
couplings and kinematical factors, for example.  The freeze-out from
thermal equilibrium occurs for $H \simeq \Gamma_\phi^\text{prod}$, which
yields the decoupling temperature
\begin{equation} \label{eq:TDcpGeneric}
 T_\phi^\text{dcp} \simeq
 \left( \frac{\pi^2 g_\ast(T_\phi^\text{dcp})}{90} \right)^\frac{1}{2}
  \frac{\Lambda^2}{x \mplanck} \simeq
 \frac{2.1 \, \Lambda^2}{x \times 10^{18}\GeV} \,.
\end{equation}

For the decay we estimate
\begin{equation}
 \Gamma_\phi^\text{dec} = y \, \frac{m_\phi^3}{\Lambda^2} \,,
\end{equation}
where $y$ contains model-dependent factors.  Generically, we expect
$x \lesssim y$, where kinematical factors and the relation between
number density and temperature tend to lead to a somewhat smaller $x$.
For instance, for the saxion we find
$x \simeq 6 \times 10^{-7}$ and $y \simeq 3 \times 10^{-6}$.
We obtain the temperature after the decay as discussed in
Sec.~\ref{sec:EntropyFromDecays},
\begin{equation} \label{eq:TDecGeneric}
 T_\phi^\text{dec} \simeq 
 \left( \frac{45}{2 \pi^2 g_\ast(T_\phi^\text{dec})} \right)^\frac{1}{4}
  \frac{( y m_\phi^3 \mplanck )^\frac{1}{2}}{\Lambda} \simeq
 1.1 \times 10^9 \,
  \frac{y^\frac{1}{2} m_\phi^\frac{3}{2} \GeV^\frac{1}{2}}{\Lambda} \,,
\end{equation}
assuming a sufficiently late decay to yield
$g_\ast(T_\phi^\text{dec}) = 10.75$.  Together with the analogon
of~\eqref{eq:tsaxeq}, which holds for any thermally produced scalar,
and~\eqref{eq:d2}, we find the dilution factor
\begin{equation} \label{eq:DeltaGeneric}
 \Delta \simeq 1.1 \times 10^{-12} \,
 \frac{\Lambda}{(y m_\phi \, \text{GeV})^\frac{1}{2}} \,,
\end{equation}
estimating as before
$\big\langle g_\ast^{1/3} \big\rangle \simeq g_\ast^{1/3}(T^\text{dec}_\phi)$.

Now we can use~\eqref{eq:o32est}, \eqref{eq:TDcpGeneric},
$\O_{3/2}^\text{tp} \leq \O_\text{DM}$
and $T_\text{R} > T_\phi^\text{dcp}$ to obtain a lower limit on $\Delta$
and thus a constraint on the model parameters,
\begin{equation}
 \Delta \gtrsim
 \frac{6.8}{x} \left( \frac{\Lambda}{10^{14}\GeV} \right)^2
 \left( \frac{M_{\widetilde g}(m_\text{Z})}{10^3 \GeV}\right)^2 
 \left( \frac{100 \GeV}{\mgrav}\right) .
\label{eq:DeltaLimitFromOmega}
\end{equation}
Furthermore,~\eqref{eq:d2}, \eqref{eq:Tafter} and~\eqref{eq:tsaxeq}
yield an upper limit on $\Delta$, which can be combined
with~\eqref{eq:DeltaLimitFromOmega}, resulting in
\begin{equation}
 \frac{\Lambda}{(xm_\phi)^\frac{1}{2}} \lesssim
 2.1 \times 10^{13}\GeV^\frac{1}{2}
 \left( \frac{10^3 \GeV}{M_{\widetilde g}(m_\text{Z})}\right)
 \left( \frac{\mgrav}{100 \GeV}\right)^\frac{1}{2} .
\end{equation}
Plugging this bound into \eqref{eq:DeltaGeneric} yields the maximal
dilution factor that can be realized with a thermally produced generic
scalar,
\begin{equation} \label{eq:DeltaMaxGeneric}
 \Delta \lesssim 24
 \left( \frac{x}{y} \right)^\frac{1}{2}
 \left( \frac{10^3 \GeV}{M_{\widetilde g}(m_\text{Z})}\right)
 \left( \frac{\mgrav}{100 \GeV}\right)^\frac{1}{2} .
\end{equation}
Using further combinations of~\eqref{eq:Tafter}
and~\eqref{eq:TDecGeneric}--\eqref{eq:DeltaLimitFromOmega}, we find that
this maximal dilution is reached for
\begin{align}
 \Lambda &\simeq 1.9 \times 10^{14} \GeV \,
 \frac{x^\frac{3}{4}}{y^\frac{1}{4}}
 \left( \frac{10^3 \GeV}{M_{\widetilde g}(m_\text{Z})}\right)^\frac{3}{2}
 \left( \frac{\mgrav}{100 \GeV}\right)^\frac{3}{4} ,
\\
 m_\phi &\simeq 79\GeV
 \left( \frac{x}{y} \right)^\frac{1}{2}
 \left( \frac{10^3 \GeV}{M_{\widetilde g}(m_\text{Z})}\right)
 \left( \frac{\mgrav}{100 \GeV}\right)^\frac{1}{2} .
\end{align}
Thus, we conclude that the generalized scenario allows for the production of
some entropy, but we do not expect a dilution factor large enough to
solve the NLSP decay problem.
In order to avoid this conclusion, we have to consider a situation where
the mechanisms for production and decay are different, so that the
decoupling temperature and the decay temperature are no longer
connected.

\subsubsection[Saxion as Oscillating Scalar]{$\Phi_\text{sax}$ as Oscillating Scalar}
As the saxion corresponds to a flat direction of the scalar potential
lifted by SUSY breaking effects, it can
develop a large field value during inflation. It begins to oscillate
around the potential minimum when the Hubble parameter becomes comparable
to the saxion mass.
This corresponds to the production of non-relativistic particles.
The temperature at the onset of oscillations is
\begin{equation}
 T^\text{osc}_\text{sax} \simeq 2.2 \times 10^{10} \GeV \left(\frac{m_\text{sax}}{1 \TeV}\right)^{\frac{1}{2}} 
 		\left(\frac{228.75}{g_\ast(T^\text{osc}_\text{sax})}\right)^{\frac{1}{4}} .
\end{equation}
Since we consider reheating temperatures higher than
$T_\text{sax}^\text{osc}$ to enable thermal leptogenesis, the produced
saxion abundance is independent of $T_\text{R}$ and given
by~\cite{Kawasaki:2007mk}
\begin{align}
 \frac{\rho^\text{osc}_\text{sax}}{s} &= \frac{1}{8} T^\text{osc}_\text{sax} \left(\frac{\phi_\text{sax}^\text{i}}{M_\text{Pl}}\right)^2 \nonumber \\
 &\simeq 4.8 \GeV \left(\frac{m_\text{sax}}{1 \TeV}\right)^{\frac{1}{2}} 
 \left(\frac{f_a}{10^{14} \GeV}\right)^2 
 \left(\frac{\phi_\text{sax}^\text{i}}{f_a}\right)^2
 \left(\frac{228.75}{g_\ast(T^\text{osc}_\text{sax})}\right)^{\frac{1}{4}},
 \label{eq:rhoosc}
\end{align}
where $\phi_\text{sax}^\text{i}$ denotes the initial amplitude of the oscillations
and where we have assumed the simplest saxion potential,
$V = \frac{1}{2} m_\text{sax}^2 \phi_\text{sax}^2$.

In this way production and decay are disconnected, circumventing the second
feature mentioned at the end of the previous section.
There is an additional free parameter, $\phi_\text{sax}^\text{i}$. 
The saxion density is constrained by requirement iv), i.e.\ that it
should not dominate before NLSP freeze-out.
 For the limiting case
of domination onset at $T^\text{fo}_\text{nlsp}$ we obtain from~\eqref{eq:help1}
\begin{equation}
\label{eq:rhoeq}
 \frac{\rho_\text{sax}^\text{osc}}{s} = \frac{3}{4} T^{=}_\text{sax} = \frac{3}{4} T^\text{fo}_\text{nlsp} \, .
\end{equation}
Equalizing~\eqref{eq:rhoosc} and~\eqref{eq:rhoeq} we find for
the initial amplitude
\begin{equation}
 \left(\frac{\phi_\text{sax}^\text{i}}{M_\text{Pl}}\right)^2 = 6 \, \frac{T^{=}_\text{sax}}{T^\text{osc}_\text{sax}}
\end{equation}
or equivalently with $T^=_\text{sax}= T^\text{fo}_\text{nlsp} \simeq m_\text{nlsp}/25$
\begin{equation}
\label{eq:sifa}
 \left(\frac{\phi_\text{sax}^\text{i}}{f_a}\right) \simeq  2.6 \times 10^4
 \left(\frac{10^{10} \GeV}{f_a}\right)   \left(\frac{8.4 \GeV}{m_\text{sax}}\right)^\frac{1}{4}
 \left(\frac{m_\text{nlsp}}{10^2 \GeV}\right)^\frac{1}{2}   \left(\frac{g_\ast(T^\text{osc}_\text{sax})}{228.75}\right)^\frac{1}{8} .
\end{equation}
The easiest expectation for the initial amplitude is
$\phi_\text{sax}^\text{i} \sim M_\text{Pl}$ or $\phi_\text{sax}^\text{i} \sim f_a$.
Interestingly, the estimate~\eqref{eq:sifa}
yields an initial amplitude $f_a < \phi_\text{sax}^\text{i} \sim \sqrt{f_a M_\text{Pl}} < M_\text{Pl}$,
if we choose the harmless value $f_a = 10^{10}\GeV$ found above.
According to~\eqref{eq:d2} and~\eqref{eq:tsaxdec}, the maximal
dilution~\eqref{eq:d3} is achieved for a saxion mass
$m_\text{sax} = 8.4 \GeV$ on the lower boundary from early enough
decay~\eqref{eq:msaxmin}.  The axion multiplet comes into thermal
equilibrium after reheating, which gives the known limit
$m_{\widetilde a} \gtrsim 1.2 \TeV$~\eqref{eq:maxinomin2}, avoiding
problems with the axino.
Thus, we have identified a working scenario where
$\rho_\text{sax}^\text{osc} \gg \rho_\text{sax}^\text{eq}$, which
enables significant entropy production while satisfying all
requirements.

Smaller $f_a$ are possible, too, provided that they respect the experimental
bound~\eqref{eq:fa}.
Larger $f_a$ and $\phi_\text{sax}^\text{i}$ were not only in conflict with
the scenario presented but also with standard cosmology.
Furthermore, larger $m_\text{sax}$ are allowed, while they lead following~\eqref{eq:tsaxdec} to smaller $\D$s.
From the naturalness point of view, the required small saxion
mass---compared to $m_{\widetilde a}$ and $m_\text{susy}$---for maximal
$\D$ might be the biggest concern.
Nevertheless, we can conclude that the saxion as oscillating scalar 
can produce
the desired entropy to soften the NLSP decay problem without violating
any constraint from cosmology or observations.

We would like to stress that our scenario does not contain more
requirements than the standard scenario with axion multiplet but no
entropy production.  Instead, we only have to change the allowed windows
for some parameters, most importantly $\phi_\text{sax}^\text{i}$ and
$m_\text{sax}$.  Avoiding axion overproduction by vacuum misalignment
even becomes easier.  Other restrictions, in particular those on $x$
in~\eqref{eq:saxtoaa} as well as on $f_a$ and $m_{\widetilde a}$, are
the same as in the standard scenario.
Note also that the initial amplitude of the saxion oscillations would
have to be restricted to small values around $f_a$, if one wanted to
prevent entropy production.  In other words, the classical field value
of the saxion endangers the standard scenario.

\section{Conclusions}

We have discussed the possibility to solve the gravitino problem in
scenarios with standard thermal leptogenesis and thus a high reheating
temperature by late-time entropy production. Our setup has been stable
gravitino dark matter with a neutralino NLSP\@.  We have estimated that thermal
leptogenesis is compatible with entropy production diluting the baryon
asymmetry as well as the LSP and NLSP relic densities by up to three to
four orders of magnitude.  This amount of dilution roughly coincides
with the maximum amount obtainable for radiation domination at NLSP
freeze-out.  

For a gravitino LSP with a mass of $100\GeV$, which allows for a
reheating temperature suitable for thermal leptogenesis, we have shown
that a neutralino NLSP which is not much heavier can be diluted
sufficiently to be compatible with BBN, i.e.\ its decays do not cause
 changes of the primordial light element
abundances that are excluded by observations.
  However, this is only possible if the lightest neutralino
contains a large wino or Higgsino component, whereas a 
bino-like neutralino remains excluded.
 
We have discussed the general requirements on the particle producing the
desired entropy and found that it is severely constrained. 
On the other hand, in some sense all these requirements either
have to be fulfilled by long-lived particles anyway or 
are generically fulfilled.
As a specific example, we have discussed the saxion
from the axion multiplet. 
We have found that sufficient entropy
production is not possible for a thermally produced saxion, where the
same couplings are relevant for its production and its decay.  This is
due to two conflicting requirements: on the one hand sufficient
production requires sufficiently strong couplings, while on the other
hand sufficiently late decay requires weak couplings,
where later decay corresponds to more entropy production.  In the
considered case, the allowed parameter ranges fail to overlap.
Using simple estimates, we have generalized this negative conclusion
to generic thermally produced particles.
Furthermore, we have encountered severe problems with the saxion's
superpartner, the axino.

As an alternative, we have considered saxion production in coherent
oscillations, which is independent of the saxion coupling.  In this
case, a relatively light saxion with a mass around $10\GeV$ is indeed
able to satisfy all requirements.  Thus, if the Peccei-Quinn mechanism solves the
strong CP problem, the potentially dangerous saxion decays can in fact
turn out as a fortune, solving the a priori unrelated gravitino problem.

\subsection*{Acknowledgements}
We would like to thank Wilfried Buchm\"uller, Laura Covi, Gudrid
Moortgat-Pick, Andreas Ringwald, Jonathan Roberts and Frank Steffen for
helpful discussions.
This work was supported by the German Science Foundation (DFG) via the
Junior Research Group ``SUSY Phenomenology'' within the Collaborative
Research Center 676 ``Particles, Strings and the Early Universe''.

\phantomsection % Ensures that a PDF bookmark is set here
\addcontentsline{toc}{chapter}{References}
\bibliography{EPbibliography}

\end{document}